\documentclass[review]{elsarticle}
\usepackage{mathtools}
\usepackage{comment}
\usepackage{float}
\usepackage{multirow}
\usepackage{amsthm}
\usepackage{graphicx}
\usepackage[export]{adjustbox}
\usepackage{changepage}
\usepackage{url}
\usepackage{amssymb}
\usepackage{threeparttable}
\usepackage[normalem]{ulem}
\usepackage{array}
\usepackage{booktabs}
\usepackage[table]{xcolor}
\usepackage{hyperref}
\usepackage{mdframed}
\usepackage{rotating}

\hypersetup
{
  colorlinks   = true, 
  urlcolor     = red,  
  linkcolor    = blue, 
  citecolor    = blue  
}
\usepackage{wasysym}
\usepackage{listings}
\usepackage{pdflscape}
\usepackage{rotating}
\usepackage[utf8]{inputenc}
\usepackage[T1]{fontenc}
\usepackage{tikz,lipsum,lmodern}
\usepackage[most]{tcolorbox}
\usepackage{pgfplots}
\usepackage[toc,page]{appendix}
\usepackage{textcomp}
\usepackage{epstopdf}
\usepackage{csquotes}
\usepackage{dirtytalk}
\usepackage{parskip}
\usepackage{amssymb}
\usepackage{tikz}
\usepackage{comment}
\usepackage{indentfirst}
\usepackage[colorinlistoftodos,prependcaption,textsize=tiny]{todonotes}
\usepackage[english]{babel}
\usepackage{geometry}
\pgfplotsset{compat=1.9}
\usepackage{lineno}
\usepackage{caption} 
\usepackage{adjustbox}
\usepackage{pdfpages}
 \usepackage{ragged2e} 
\usepackage{xurl}
\usepackage{longtable,tabu}
\usepackage{afterpage}
\usepackage{verbatim}
\usepackage{subfigure}
\usepackage{subfloat}
\usepackage{longtable}
\usepackage{ltxtable}
\usepackage{fancyvrb}
\usepackage{tabularx}
\usepackage{makecell}
\usepackage{supertabular}

\journal{Elsevier}

\begin{document}
\begin{frontmatter}

\title{UML Class Diagram Evaluation and Repair Strategies based on LLMs}
    

\author[mymainaddress]{Jie Liang}
\ead{liangjie@wsyu.edu.cn}
\address[mymainaddress]{School of Information Science and Engineering, Wuchang Shouyi University, Wuhan, China}

\author[mysecondaryaddress]{Peng Liang\corref{mycorrespondingauthor}}
\cortext[mycorrespondingauthor]{Corresponding author at: School of Computer Science, Wuhan University, China. Tel.: +86 27 68776137; fax: +86 27 68776027.}
\ead{liangp@whu.edu.cn}


\author[mysecondaryaddress]{Chong Wang\corref{mycorrespondingauthor}}
\ead{cwang@whu.edu.cn}
\address[mysecondaryaddress]{School of Computer Science, Wuhan University, Wuhan, China}

\begin{abstract}
UML class diagrams are a crucial tool for defining the structure of software systems, but designing accurate and comprehensive class diagrams is a challenging task. Traditionally, creating UML models relies on the expertise and experience of professionals. However, with the development of AI technologies, particularly large language models (LLMs), new opportunities for software modeling have emerged. Despite this, there has been limited research on the application of LLMs in software modeling, especially in UML class diagram modeling. This study adopts an exploratory research paradigm and conducts empirical experiments on several typical software system cases. Combining the SDMetrics measurement tool with expert manual review, this paper comprehensively evaluates the practical performance of mainstream LLMs in UML class diagram modeling from multiple dimensions, including  size and completeness, relationship correctness, inheritance hierarchy, and design rule compliance. Focusing on typical defects in LLM‑generated UML class diagrams, this study reveals that LLMs exhibit uncertainties analogous to human memory. Accordingly, three targeted repair strategies are proposed, including memory reinforcement, external knowledge injection, and detection-guided automated targeted repair. Experimental results obtained from the case studies indicate that (1) compared to expert‑crafted class diagrams, LLM-generated UML class diagrams exhibit several issues, such as incomplete identification of key classes, confusion or omissions in relationships, insufficient or absent inheritance relationships, unused classes, and circular dependencies, and (2) after applying the repair methods, all the LLMs show varying degrees of improvement in addressing these issues. The average repair rate for key class identification reaches 85\%, the coupling relationship repair rate is 46\%, the inheritance relationship repair rate is 69\%, while repair rates for unused classes and circular dependencies both reach 100\%. This study highlights both the practical limitations and optimization potential of LLM-based UML modeling, offering empirical insights and actionable strategies for advancing intelligent software modeling research.
\end{abstract}



\end{frontmatter}

\section{Introduction}\label{chap:intro}
Software modeling is a critical component of software engineering and directly affects the performance and reliability of the final software product~\citep{camara2024towards}. UML class diagrams, as an important tool for defining software system architecture, visually represent classes, their attributes, methods, and relationships~\citep{camara2024towards}. They serve as a crucial bridge for transforming domain knowledge into executable designs. However, constructing high-quality class diagrams poses two major challenges. On the one hand, requirements analysis requires abstracting ambiguous business descriptions into structured models that comply with UML semantics, demanding both domain insight and modeling expertise from practitioners~\citep{hou2024large}. On the other hand, traditional approaches that rely on expert manual design suffer from efficiency bottlenecks and struggle to accommodate the rapid iterations required by agile development~\citep{hou2024large}.

Recent breakthroughs in large language models (LLMs) offer new opportunities to transform software modeling paradigms~\citep{fan2023large}. LLMs, such as ChatGPT and DeepSeek, demonstrate exceptional capabilities in natural language understanding and generation, and are increasingly being applied to software engineering tasks such as code generation and requirements analysis~\citep{fan2023large,ahmad2023human}. However, research on applying LLMs to the core modeling task of UML class diagram generation remains nascent. Most existing work is limited to preliminary performance validation of individual models and lacks systematic evaluation frameworks~\citep{marques2024chatgpt}. Moreover, effective quality enhancement strategies have yet to emerge for addressing common structural defects in generated class diagrams, such as redundant associations and mixed responsibilities~\citep{naveed2024mde}.

To this end, this study pursues a primary objective: to evaluate and enhance the quality of UML class diagrams generated by large language models (LLMs). To achieve this objective, three cases were selected to represent diverse application domains and varying levels of complexity. For each case, expert‑crafted class diagrams serve as the validated gold standard for assessment. Subsequently, three representative LLMs with distinct capabilities were selected for experimentation: OpenAI's general‑purpose model GPT‑4, OpenAI's reasoning model o1‑preview, and DeepSeek's reasoning model DeepSeek‑R1. Taking these expert‑crafted class diagrams as the gold standard, this study develops a comprehensive assessment framework to enable multidimensional and precise evaluation of LLM‑generated class diagrams. The framework encompasses design size, coupling degree, inheritance hierarchy, and compliance with design rules~\citep{disipio2024llms,siala2025using}. A human‑machine collaborative approach is adopted for systematic analysis. Finally, to address key quality issues in LLM‑generated class diagrams, this study proposes several improvement strategies, including external knowledge injection, memory reinforcement, and automated defect detection with targeted repair, and empirically evaluates their effectiveness.

Experimental results demonstrate that LLM-generated UML class diagrams consistently suffer from recurring modeling defects compared with expert-crafted benchmarks, including inaccurate key class identification, erroneous or missing class relationships, incomplete inheritance structures, unused classes, and cyclic dependencies. Furthermore, the proposed repair strategies can effectively alleviate these defects across different LLMs and achieve substantial improvements in overall diagram quality.

The \textbf{main contributions} of our work are as follows:
\begin{itemize}
    \item We propose a comprehensive method for evaluating the quality of LLM-generated class diagrams. The method compares expert‑crafted class diagrams with LLM-generated outputs across key dimensions, including design size, coupling, inheritance hierarchy, and compliance with design rules. By combining automated analysis using SDMetrics~\citep{wust2005sdmetrics} with manual review, the method enables an in-depth assessment of the quality of generated class diagrams.
    \item We propose a set of repair strategies and evaluation schemes to address quality defects in LLM-generated class diagrams. Specifically, we introduce external knowledge injection and targeted repair based on the results of automated defect detection. In addition, motivated by the similarity between uncertainty in large language models and the forgetting characteristics of human memory, we propose a memory-reinforced repair strategy.
\end{itemize}

\textbf{Paper Organization}: Section \ref{RelatedWork} surveys existing research concerning LLM‑based UML modeling, quality assessment, and defect repair. It summarizes representative empirical studies and highlights two major research gaps: the absence of multidimensional evaluation frameworks for LLM‑generated class diagrams and the lack of systematic strategies for addressing structural defects. Section \ref{chap:case} presents the research methodology. The research design is organized into four sequential stages: data collection, class‑diagram generation, evaluation of LLM‑generated class diagrams, and class‑diagram repair. Data collection prepares case resources and expert‑crafted reference class diagrams. We then generate UML class diagrams via different LLMs, evaluate their modeling quality, and implement targeted repairs for detected defects. Section \ref{sec:results} presents the two research questions and their corresponding results. Section \ref{sec:discussion} discusses the findings and their implications, while Section \ref{sec:Threats} addresses the limitations of this study. Finally, Section \ref{conclusions} presents the conclusions and directions for future work. 

\section{Related Work}\label{RelatedWork}\label{chap:relat}
Large language models (LLMs) are rapidly advancing the field of software engineering~\citep{jiang2025survey}. Their development has introduced new approaches and methodologies for UML class diagram modeling while also drawing scholarly attention to two critical questions. First, how can the quality of LLM-generated class diagrams be evaluated in a rigorous and comprehensive manner to clarify their strengths and limitations? Second, how can issues in the generated diagrams be identified and effectively addressed to improve the resulting models~\citep{jiang2025survey,wang2024llms}.

The application of LLMs to software engineering has attracted increasing attention, and numerous studies have investigated their performance in UML modeling. Wang et al.~\citep{wang2024llms} engaged 45 undergraduate software engineering students in UML modeling tasks involving use case, class, and sequence diagrams with LLM assistance. Through a manual analysis of the students' project reports, they examined the models' ability to identify relevant modeling elements across different UML diagram types, as well as the influence of prescribed output formats on modeling performance. Cámara et al.~\citep{camara2023assessment} evaluated ChatGPT's performance in software modeling through a series of experiments that assessed its ability to generate syntactically and semantically correct UML models, adapt to different contexts and problem domains, handle varying model sizes, effectively employ modeling concepts and mechanisms, respond to prompt variations, and adhere to UML notation. They further proposed a conceptual framework for the standardized benchmarking of LLMs in software modeling tasks~\citep{camara2024towards}. The framework combines manually specified requirements with two phases of automated execution and analysis to evaluate the quality of generated software models. De Bari et al.~\citep{debari2024evaluating} assessed the ability of LLMs to generate UML class diagrams by comparing their outputs with human-produced solutions. Their evaluation considered syntactic, semantic, and pragmatic correctness, as well as the distance between the generated and reference solutions. Mina et al.~\citep{shehata2024creating} investigated the reverse-generation task of deriving class diagrams from source code. Using the LindHolmen dataset, which comprises 40 open-source projects, they systematically evaluated ChatGPT's accuracy and structural soundness through refined prompt-engineering strategies, including PlantUML format constraints and code-input integrity checks.

These pioneering studies laid the foundation for research on the use of LLMs in UML modeling. However, existing work has predominantly pursued single-dimensional investigations and has yet to establish a comprehensive framework for evaluating class diagram quality~\citep{joel2024survey}. Because diagram quality directly affects system characteristics, current research lacks a multidimensional and in-depth analysis of structural defects in LLM-generated diagrams~\citep{joel2024survey,zhang2024systematic}. Motivated by the emphasis of prior work on the correctness of modeling elements, this paper develops a multidimensional evaluation system. By combining automated tool-based analysis with manual review, the proposed approach enables a more systematic assessment of the structural soundness of class diagrams and provides a comprehensive analytical perspective for future research.

Research on improving the quality of LLM-generated UML models has primarily focused on task decomposition and the exploration of generation strategies. Chen et al.~\citep{chen2024model} proposed a decomposition-based modeling approach that divides class diagram generation into three independent stages to improve the completeness and semantic consistency of diagram elements. More recently, the NOMAD framework, inspired by cognitive science, introduced a multi-agent pipeline that decomposes modeling tasks into role-specific subtasks and has achieved promising results on benchmark datasets. It also introduced the first error-classification scheme for LLM-generated UML class diagrams, although limitations remain in attribute-extraction accuracy and agent coordination. Ferrari et al.~\citep{ferrari2024model} proposed optimization strategies, including requirements-quality validation, for UML sequence diagrams generated by GPT-3.5. However, the details of their experimental design remain unclear, and the effectiveness of these strategies was not validated through controlled experiments. Existing research on UML model repair has largely focused on inducing error types, with limited attention to systematic repair strategies~\citep{feng2025integrating}. Although NOMAD's validation agents can correct certain surface-level errors, they struggle to address hierarchical defects in the generated models~\citep{giannouris2026nomad}. 

Although prior research has demonstrated the potential to improve UML model quality, several key practical challenges remain insufficiently addressed~\citep{jain2024livecodebench}. In particular, existing repair-oriented studies have focused primarily on error induction and surface-level error correction, whereas systematic repair strategies for addressing structural defects in LLM-generated class diagrams remain largely unexplored~\citep{feng2025integrating,giannouris2026nomad}. To address this gap, this paper proposes a set of repair strategies for mitigating such structural defects and offers new insights into improving UML model quality. Further details of the research design are presented in Section~\ref{chap:case}.

\section{Research Design}\label{chap:case}

\begin{figure}[!t]
\centering
\includegraphics[width=1\textwidth]{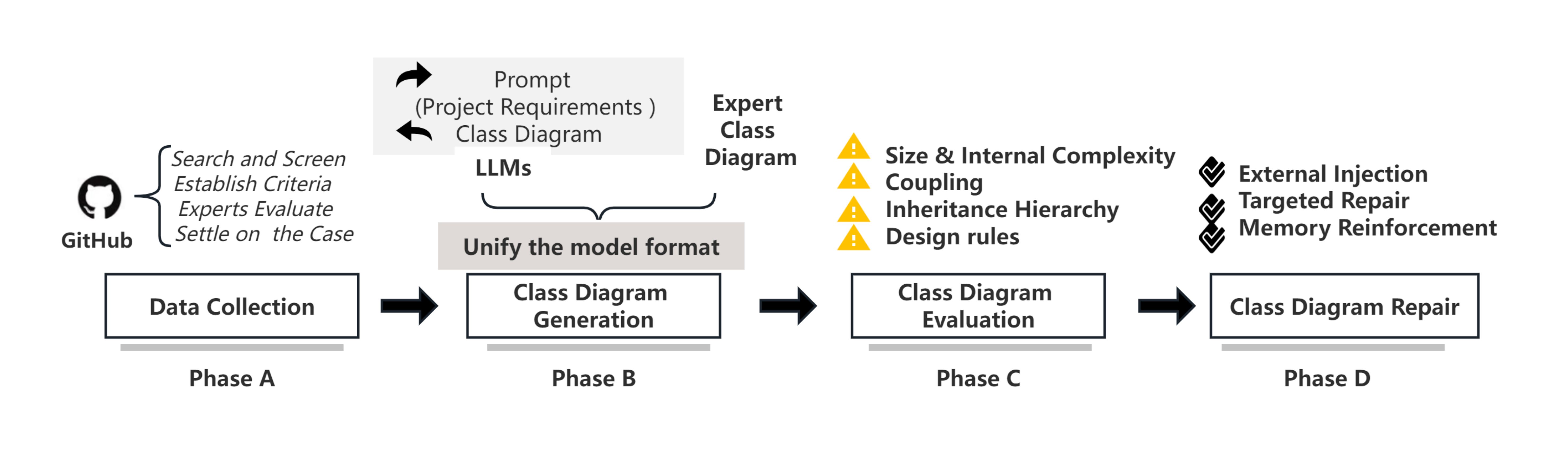}
\caption{Overview of the research process}
\label{fig:Overview}
\end{figure}

The objective of this study is to evaluate the quality of LLM-generated class diagrams and repair the identified defects. To achieve this objective, we designed and conducted a series of experiments, the overall research process of which is illustrated in Figure~\ref{fig:Overview}.

\subsection{Data Collection}\label{sec:data}
The first author conducted an extensive search on GitHub for software design case studies. The search terms covered a wide range of common application domains, including e-commerce, finance, healthcare, and education, as well as various software design patterns and technology stacks, thereby ensuring the diversity of the collected cases. The identified cases were then subjected to preliminary screening, with priority given to those that explicitly included requirements documents, such as requirements specifications and user stories, together with detailed design artifacts, including class diagrams and supporting design documentation.

To ensure the scientific validity of both the case studies and the expert-crafted class diagrams, we established three evaluation criteria: class diagram complexity, class diagram quality, and requirements readability. Regarding complexity, each class diagram was required to contain at least 10 entity classes, with each class having at least 2--3 attributes. The quality of the expert-crafted class diagrams was assessed using a five-point Likert size~\citep{kusmaryono2022number}. The three authors independently evaluated the quality of each diagram, and their scores were averaged. The quality assessment comprised four components: the alignment of class definitions with business requirements (weight: 0.3), the extent to which class attributes and methods reflected the characteristics and behaviours of their respective classes (weight: 0.3), the conformity of class relationships to the underlying business logic (weight: 0.2), and the readability and extensibility of the class diagram structure (weight: 0.2). An average score of at least 4.5 was required. If the difference between the scores assigned by any two authors for a given expert-crafted class diagram exceeded 0.5 points, the third author was consulted, and the authors jointly discussed the discrepancy to determine the final score. The requirements-readability score for each project was required to exceed 70, indicating relatively easy readability. We used the online text-readability analysis tool TextGears (textgears.com) to assess the readability of the project requirements.

Following the screening and evaluation process, three projects from distinct domains were selected as experimental subjects: Parking Lot (an IoT scenario)~\citep{manualdesigns2025parking}, Online Movie (an e-commerce scenario)~\citep{manualdesigns2025movie}, and Stack Overflow (a knowledge-service scenario)~\citep{manualdesigns2025stack}. Detailed assessment information for each project, including natural-language requirements descriptions and expert-crafted class diagrams, is presented in Table~\ref{tab:project_evaluation}.

\begin{table}[htbp]
  \centering
  \caption{Project Evaluation Metrics and Results}
  \label{tab:project_evaluation} 
  \resizebox{\textwidth}{!}{
    \begin{tabular}{lcccccc}
      \toprule
      \multirow{2}{*}{Project Name} & \multicolumn{3}{c}{Class Diagram Complexity} & \multirow{2}{*}{Class Diagram Quality} & \multirow{2}{*}{\makecell[c]{Requirement Readability\\(Description)}} \\
      \cmidrule{2-4} 
      & Number of Classes & Number of Methods + Attributes & Number of Relationships & & \\
      \midrule
      Parking Lot  & 28 & 57 & 30 & 4.6 & 81.1 (Easy to read) \\
      Online Movie   & 22 & 66 & 26 & 4.7 & 73.2 (Relatively easy to read) \\
      Stack Overflow  & 13 & 54 & 19 & 4.7 & 75.1 (Relatively easy to read) \\
      \bottomrule
    \end{tabular}
  }
\vspace{2pt} 
\end{table}

Table~\ref{tab:project_evaluation} presents the evaluation results for the three cross-domain projects selected for this study: Parking Lot, Online Movie, and Stack Overflow. The UML class diagrams were assessed along three key dimensions: diagram complexity, measured in terms of the numbers of classes, attributes and methods, and relationships; diagram quality; and requirements readability. The results show a gradual variation in project complexity, with the number of classes ranging from 13 to 28. Meanwhile, the class diagram quality scores remained consistently high, at no less than 4.6. The requirements-readability scores ranged from 73.2 to 81.1.

\subsection{Class Diagram Generation}\label{sec:criteria}
After the project cases were identified, the class diagram creation process was conducted in two stages. First, the natural-language requirements of each case were used as prompts for LLMs to generate class diagrams automatically. Second, the formats of the LLM-generated and expert-crafted class diagrams were standardized to enable their subsequent evaluation using the SDMetrics tool~\citep{wust2005sdmetrics}.

\subsubsection{LLM-generated class diagrams}\label{LLM-generated class diagrams}
To thoroughly explore the modeling capabilities of large language models and comprehensively compare differences across versions, we selected three widely used models. These include two models from OpenAI: GPT-4.0 and o1-preview, along with DeepSeek-R1, an inference model developed by DeepSeek Technology. GPT-4.0 was released on March 14, 2023, and was the leading language model at the time. In September 2024, OpenAI officially launched its latest AI reasoning model series—o1-preview. This series focuses on simulating human reasoning processes, delivering more precise answers through deep contemplation and multidimensional strategies. In January 2025, DeepSeek released its reasoning model DeepSeek-R1. This model employs reinforcement learning for post-training, aiming to enhance reasoning capabilities, particularly excelling at complex tasks such as mathematics, programming, and natural language reasoning.DeepSeek-R1 demonstrates outstanding reasoning capabilities in specialized domains such as medical clinical decision-making, yet its performance falls short in complex UML-driven coding scenarios~\citep{li2025prompting}. Research on its capabilities in software modeling tasks remains relatively limited at present.

To ensure the objectivity and impartiality of the experimental results, this study adopts a one-shot generation paradigm for UML class diagram construction rather than a multi-round generation and optimal-sample-screening procedure. Specifically, standardized project-requirement prompts are submitted to the three LLM platforms under consistent conditions. Each model is instructed to generate a UML class diagram that strictly conforms to the specified requirements and to provide the result in a standardized PlantUML format. The generated PlantUML code is then rendered directly using the PlantUML online platform to obtain each model's original UML class diagram output.

Unlike the repeated-generation and manual-selection procedures commonly used in existing studies, this study does not involve repeated generation, manual filtering, or the selection of superior model outputs. Each test case therefore corresponds to one independent one-shot generation from each LLM, preserving the original output characteristics and inherent stochasticity of the models under standard inference conditions. This design avoids the upper-bound performance bias introduced by manually screening for optimal samples and eliminates potential circularity between the screening criteria and the evaluation metrics. All original outputs are included directly in the subsequent multidimensional evaluation, ensuring that the experimental data accurately reflect the performance of mainstream LLMs under conventional single-call inference in UML modeling tasks. Consequently, the evaluation results are more objective, fair, and reproducible.

\subsubsection{Unify the Model Format}\label{Unify the Model Fomat}
After the class diagrams are generated, their formats are standardized using the Visual Paradigm modeling tool. Both the expert-crafted and LLM-generated class diagrams are formatted consistently and exported in XML Metadata Interchange (XMI) format. The resulting XMI files are then imported into SDMetrics for subsequent quality assessment.

\subsection{Class Diagram Evaluation}\label{Evaluation of Class Diagrams}
We evaluate the expert-crafted and LLM-generated class diagrams across four dimensions: size, relationships, complexity, and adherence to design rules. The evaluation combines automated testing tools with manual review.

\subsubsection{Size and Internal Complexity Metrics}\label{Scale}
Number of Classes. The number of classes provides an indication of a system's size and Internal Complexity . An excessive number of classes may suggest an overly fragmented design, resulting in complex interactions among system components and increased difficulty in comprehension and maintenance. Conversely, too few classes may indicate insufficient functional decomposition, potentially causing individual classes to assume excessive responsibilities.

Analysis Method: The number of classes is obtained from the metric data generated by SDMetrics. Its appropriateness is assessed in relation to the system's functional requirements and expected size.

Number of Class Attributes and Methods. An excessive number of attributes or methods within a class may indicate high internal complexity. Too many attributes can make the class's state difficult to manage, whereas too many methods may suggest that the class performs a wide range of unrelated operations. Both conditions may indicate a violation of the Single Responsibility Principle, thereby reducing the maintainability and comprehensibility of the class.

Analysis Method: The numbers of attributes and methods in each class are examined using the SDMetrics data. Empirical thresholds are applied as preliminary indicators; for example, a class with more than 10 attributes or 20 methods is subject to closer examination. The necessity and coherence of these attributes and methods are then evaluated in relation to the class's intended responsibilities.

\subsubsection{Coupling Metrics}\label{Relationships}
The number of inter-class relationships reflects the degree of coupling within a system. This coupling is characterized by four types of relationships: dependency, association, aggregation, and composition. A high degree of coupling creates strong interdependencies among classes, such that modifications to one class may affect multiple others and require corresponding changes to their dependents. These cascading effects can increase maintenance costs and introduce additional risks throughout the system.

Analysis Method: The coupling of each class is assessed by examining its incoming dependencies (the classes that depend on it) and outgoing dependencies (the classes on which it depends). SDMetrics provides the following six metrics for quantifying inter-class coupling:

IC\_Attr: Number of attributes in a class whose types are other classes or interfaces

EC\_Attr: Number of external attributes for which the class serves as the attribute type

IC\_Par: Number of parameters in a class whose types are other classes or interfaces

EC\_Par: Number of external parameters for which the class serves as the parameter type

Dep\_Out: Number of elements, such as other classes or interfaces, on which the class depends

Dep\_In: Number of elements, such as other classes or interfaces, that depend on the class

For each class, the values of EC\_Attr, EC\_Par, IC\_Attr, IC\_Par, Dep\_Out, and Dep\_In are summed to obtain a composite coupling metric. A higher value indicates a greater degree of inter-class coupling~\citep{briand1999unified}.Classes whose coupling metric values are substantially higher than those of other classes are identified as potentially highly coupled elements and therefore require particular attention during design and maintenance. The validity of their relationships is further assessed through manual review, comparison with system requirements, and evaluation against established design principles.

\subsubsection{Inheritance Hierarchy Complexity Metrics}\label{Complexity}
The structural complexity of a class diagram can be assessed through the depth of its inheritance hierarchy. An excessively deep hierarchy may hinder comprehension and maintenance because subclasses inherit attributes and operations through multiple levels. Consequently, modifications to an ancestor class may propagate to numerous descendants.

Analysis Method: Inheritance-depth values are obtained from the SDMetrics results. If an inheritance hierarchy exceeds three to five levels---an empirical range that may be adjusted according to the system context---its design is subjected to further examination. SDMetrics provides two metrics for assessing inheritance depth: DIT (Depth in Inheritance Tree), which measures the distance from a class to the root of the hierarchy, and CLD (Class-to-Leaf Depth), which measures the maximum distance from a class to a descendant leaf.

The combinations of these metrics provide complementary insights into class-hierarchy complexity~\citep{fenton2014software}. High DIT and high CLD values indicate that a class is deeply nested while retaining a substantial hierarchy beneath it, suggesting a complex inheritance structure in which changes may have multilevel effects. High DIT and low CLD values characterize classes located deep in the hierarchy but close to leaf nodes; although their descendant scope is limited, they may still be affected by changes propagated through multiple ancestors. Conversely, low DIT and high CLD values identify classes near the root that have long descendant paths, making modifications potentially consequential for many levels of subclasses. Low DIT and low CLD values indicate a comparatively shallow and compact inheritance structure that is generally easier to understand and maintain.

The joint analysis of DIT and CLD therefore provides a systematic means of evaluating inheritance-hierarchy complexity and supports decisions concerning system design, optimization, and maintenance~\citep{shaikh2021more}.

\subsubsection{Design Rule Metrics}\label{Design Rules}
Rule-based analysis can automatically identify incomplete, incorrect, redundant, or inconsistent elements in class-diagram designs. Such defects directly affect both model quality and system reliability. Incomplete designs may omit required functionality, incorrect designs may lead to implementation or runtime errors, redundant elements increase structural complexity and maintenance costs, and inconsistencies reduce the comprehensibility and maintainability of the system.

Analysis Method: Design-rule violations are identified using the rule checker. Each reported violation is then examined to determine its underlying cause and potential impact on the system.

Three common violations detected by the class-diagram rule checker are described below:

Unused (class not used elsewhere): The message ``Completeness 1-high: the class is not used anywhere'' indicates that a class does not participate in any detected relationship or usage within the system. Such classes introduce unnecessary model and codebase complexity and may impede subsequent maintenance and refactoring activities~\citep{debari2024evaluating}.

DepCycle (circular dependency): The message ``Style 2-Med: the class has circular reference'' indicates the presence of a cyclic dependency among classes. Circular dependencies may complicate object lifecycle and memory management~\citep{briand1999unified}, consume additional resources in long-running systems, degrade performance, and increase the cost and risk of modifying the class structure~\citep{shaikh2021more}.

AttrNameOvr (attribute override): The warning ``Naming 2-Med: Attribute Override'' indicates that a class declares an attribute with the same name as an inherited attribute. The new declaration may obscure the inherited attribute, undermine semantic consistency, and cause deviations in the intended business logic. Because this issue can be difficult to detect during routine inspection, it may substantially increase maintenance effort~\citep{debari2024evaluating}.

\subsection{Class Diagram Repair}\label{gptprompt}
To address the deficiencies revealed by the metric-based and qualitative analyses of automatically generated class diagrams, a series of targeted repair methods is proposed. These methods are intended to improve the accuracy and completeness of the generated diagrams, thereby supporting the demand for high-quality modeling in software engineering.

\subsubsection{External Knowledge Injection Learning Repair}\label{External Knowledge Injection Learning Repair}
This method employs external knowledge injection to repair automatically generated class diagrams. By incorporating knowledge from software engineering and other relevant professional domains, it enables the model to better understand established modeling principles and generate class diagrams that conform to domain standards~\citep{ren2024misuse}.

Implementation Details: First, diverse knowledge resources concerning UML class diagrams are collected from the software engineering domain. Essential knowledge elements, including concepts, principles, design patterns, and representative examples, are then extracted from the selected resources and organized into a structured knowledge repository. Next, this knowledge is transformed into a format suitable for model input and incorporated into the training process, allowing the model to acquire domain-specific concepts, principles, and standards. Drawing on the injected knowledge, the model performs a comprehensive analysis of problematic UML class diagrams and applies appropriate repairs to the identified defects. Finally, an evaluation mechanism is established to assess the quality of the repaired diagrams. The prompt used to repair class diagrams generated by large language models through external knowledge injection is presented in the gray box below.

\begin{tcolorbox}[breakable]
\textbf{Prompts for External Knowledge Injection Learning Repair}:
\\
\textbf{Prompt 1.1 ($P_{\mathrm{Cat}}$): the Categorization prompt.} Learn to identify entity classes and objects: Most objects can be categorized into
five types. Perceptible physical entities, such as cars, books, credit cards. Roles of people or
organizations, such as students, teachers, managers, administrators, supply offices. Events requiring
memory, such as withdrawing cash, flying, placing an order. Interactions between two or more
objects, such as purchasing, marrying. Concepts requiring explanation, such as insurance
policies, business rules.\\

\textbf{Prompt 1.2 ($P_{\mathrm{Ident}}$): the Identification prompt.} Based on the methods you learned, identify all entity class objects required by the system according to the following system requirements: \textcolor{blue}{System Requirements}\\

\textbf{Prompt 1.3 ($P_{\mathrm{Gen}}$): the Generation prompt.} Create a UML class diagram based on these entity class objects and output it in PlantUML format.
\end{tcolorbox}

\subsubsection{Targeted Repairs Based on Design Rule Checking Results}\label{Targeted Repairs}
Implementation Details: First, SDMetrics parses the automatically generated class diagram into an internal data structure suitable for analysis. It then systematically evaluates each model element and relationship against the rules in its built-in library. The detected violations are classified and prioritized according to their potential effects on the quality and functionality of the diagram~\citep{sdmetricsmanual}. Based on these findings, a targeted remediation strategy is formulated and applied to the generated model. The repair process may involve revising model parameters, correcting the definitions of class-diagram elements, or modifying the relationships among them. After the repairs have been completed, the revised class diagram is verified to determine whether the identified errors have been effectively corrected~\citep{wille2017automated}. SDMetrics then re-evaluates the repaired diagram, and the results obtained before and after repair are compared to confirm that the detected violations have been appropriately addressed. The prompt used to repair class diagrams generated by large language models through targeted remediation is presented in the gray box below.

\begin{tcolorbox}[breakable]
\textbf{Prompt for Targeted Repair}:
\\
\textbf{Prompt 2.1 ($P_{\mathrm{TgtRep}}$): the Targeted‑Repair prompt.} The ParkingAttendant class in the diagram below has the issue “the class is not used anywhere”. Please repair this issue. The diagram is: \textcolor{blue}{Unrepaired Class Diagram in PlantUML}

\end{tcolorbox}

\subsubsection{Memory Reinforcement and Repair}\label{Memory Reinforcement}
This method draws on principles of human memory by using reinforcement to improve the retention of critical information and reduce uncertainty in software modeling.

Human memory is inherently uncertain and multidimensional because of both the properties of the memory system and the influence of external factors. Rather than recording every detail as a video recorder would, human memory reconstructs information in the brain. Memories may gradually weaken or disappear over time, and they may interfere with one another, particularly when they are formed close together or contain similar information. Moreover, memory systems vary across individuals as a result of differences in genetics, environment, education, and other factors. Consequently, people may recall the same information or event differently, making memory behavior difficult to predict and interpret and increasing uncertainty in human memory~\citep{schacter1999seven,richards2023persistence}. Large language models also exhibit uncertainty in their data repositories. This uncertainty may arise from heterogeneous data sources, inconsistent data quality, annotation errors, distributional biases, and outdated information. These factors can impair the model's language understanding and generation capabilities, potentially affecting downstream applications~\citep{zhou2024bias}.

In human memory, the repeated reinforcement of accurate information can deepen its processing and storage, thereby forming more robust memory traces. This process may reduce the influence of erroneous information and distractions, preserve the clarity and accuracy of memories, and mitigate biases or errors during recall~\citep{richards2023persistence}. A similar principle can be applied to the training of large language models, where the reinforcement of accurate information derived from the model's existing knowledge can be interpreted as a form of training-data optimization. By filtering and selecting high-quality textual data, the model can learn more precise and reliable linguistic patterns and domain knowledge. Reinforcing correct information can also reduce noise and errors in the data repository, improving the model's accuracy and reliability in subsequent applications.

Implementation Methodology: First, a set of Chain-of-Thought (CoT)-based prompts is designed with reference to iterative bootstrapping mechanisms~\citep{sun2024iterative}. These prompts encourage the large language model to retrieve UML class-diagram-related information from its existing knowledge base. The information generated in response to these prompts is then collected into a candidate memory repository. Experienced professionals, such as UML experts and senior software engineers, manually review the repository according to predefined criteria for accuracy and completeness. Following the exemplar-selection and iterative-refinement strategy of Iter-CoT~\citep{sun2024iterative}, the verified information is corrected, organized, and formatted as training data suitable for the model. The model is subsequently instructed to repair problematic UML class diagrams using this verified information; the verified information is used as external guidance and is not reinforced by requiring the model to generate diagrams from memory. Finally, the repaired UML class diagrams are evaluated to measure the accuracy and effectiveness of the proposed repair method.

\begin{tcolorbox}[breakable]
\textbf{Prompts for Memory Reinforcement Repair(Repairing Coupling Relationships) }:
\\
\textbf{Prompt 3.1 ($P_{\mathrm{Def}}$): the Definition prompt.} What are the \textbf{fundamental definitions} of various relationships in UML class models? For example, the meanings of dependency relationships, unidirectional associations, bidirectional associations, composition relationships, and aggregation relationships.\\

\textbf{Prompt 3.2 ($P_{\mathrm{Repr}}$): the Representation prompt.} Consider how these relationships differ in representation—such as connection styles and arrow directions in class diagrams.\\

\textbf{Prompt 3.3 ($P_{\mathrm{Sem}}$): the Semantics prompt.} Analyze the semantic distinctions these relationships convey in practical application scenarios.\\

\textbf{Prompt 3.4 ($P_{\mathrm{Rev}}$): the Review prompt.} Review the methods for distinguishing UML class model relationships derived earlier, including definitions, representation styles, and semantic differences.\\

\textbf{Prompt 3.5 ($P_{\mathrm{Anal}}$): the Analysis prompt.} Carefully analyze the responsibilities and functions of each class in the existing UML class model, along with their current interaction patterns: \textcolor{blue}{Unrepaired class diagram PlantUML}  \\
\\
\textbf{Prompt 3.6 ($P_{\mathrm{Redes}}$): the Redesign prompt.} Based on the differentiation method and analysis of the existing class model, consider how to rearrange the connections between classes and adjust arrow directions to accurately reflect their intended relationships. Please redesign the relationships between classes in the above UML class model.
\\
\\
\textbf{Prompts for Memory Reinforcement Repair (Repairing Inheritance Relationships) }:
\\
\textbf{Prompt 4.1 ($P_{\mathrm{InhDef}}$): the Inheritance‑Definition prompt.} What constitutes an inheritance relationship between classes in a UML class model? What benefits does designing class inheritances bring to a system?\\

\textbf{Prompt 4.2 ($P_{\mathrm{Metr}}$): the MetricAwareness prompt.} What do the DIT and CLD parameters in a UML class model signify? What hazards does DIT>2 or CLD>2 pose to a system?\\

\textbf{Prompt 4.3 ($P_{\mathrm{Redes}}$): the Redesign prompt.} Carefully examine the unresolved UML class diagram below (presented in PlantUML format). Analyze the issues and inconsistencies in the current inheritance relationships. Based on your prior learning about inheritance in UML class models, redesign the inheritance relationships for this diagram: \textcolor{blue}{Unrepaired class diagram PlantUML} 
\end{tcolorbox}

\section{Results}\label{sec:results}
This section presents the experimental results and addresses the following two research questions (RQs).

\textbf{RQ1}: \textit{How does the overall performance of LLMs compare with that of software engineering experts in UML class-diagram modeling tasks?} This research question investigates the differences between class diagrams generated by LLMs and those created by experts. The diagrams are evaluated using SDMetrics and manual inspection across four dimensions: size and completeness, relationship correctness, inheritance hierarchy, and design rule compliance.

\textbf{RQ2}: \textit{To what extent can the proposed repair strategies improve UML class diagrams generated by LLMs?} This research question assesses the effectiveness of memory reinforcement, external knowledge injection, and detection-driven targeted repair in mitigating common defects, including missing critical classes, incorrect or omitted relationships, inadequate or absent inheritance structures, unused classes, and circular dependencies.

\subsection{Answer to RQ1}\label{RQ1}
This section systematically compares the performance of multiple LLMs with that of software engineering experts in UML class diagram modeling to address RQ1. We evaluate the quality of the LLM-generated class diagrams across four dimensions—size and completeness, relationship correctness, inheritance hierarchy, and design rule compliance—by combining quantitative metrics derived from SDMetrics with human evaluation.

\subsubsection{Size and Completeness Analysis}\label{competency4UC}
To assess whether the size of a class diagram is reasonable, we first use SDMetrics to count the number of classes and then evaluate the results based on the requirements documents and use-case analyses. Following the classification of classes in \citep{ibanez2025multimodal}, we classify classes into three categories according to their relevance to the requirements:

\begin{itemize}
    \item \textbf{Core Class.} Classes that directly correspond to explicitly mentioned core business entities or functional modules in the requirements and support the primary system logic. Examples include ``Movie'' and ``Parking Lot''.
    \item \textbf{Implicit Class.} Classes that are not explicitly stated in the requirements but are derived from implicit business rules, extended scenarios, or domain knowledge to ensure the completeness of the business logic. Examples include ``Movie Query'' and ``State Transporter''.
    \item \textbf{Management Class.} Classes that are not explicitly covered in the requirements but are introduced to support system use, operation and maintenance, or non-functional requirements, such as permission control, logging, and exception handling. Examples include ``Administrator'' and ``Operation Audit Class.'' These classes enhance system maintainability and stability.
\end{itemize}

This classification enables a systematic evaluation of whether a class diagram comprehensively covers the requirements without omitting key classes, appropriately captures potential business logic through the inclusion of implicit classes, and satisfies non-functional system requirements through the inclusion of necessary management classes. Together, these criteria determine whether the size of the class diagram is appropriate.

Table \ref{tab:llm_expert_class_coverage_table2} presents a comparison of class coverage between software engineering experts and LLMs in class diagram generation. The main findings are as follows:

\begin{itemize}
    \item \textbf{Core Classes.} Across the three projects, there are 47 core classes in total. GPT-4 covered 28 (59.6\%), o1-preview covered 44 (93.6\%), and DeepSeek-R1 covered 36 (76.6\%). For example, in the Parking Lot scenario, GPT-4 achieved only 48\% core-class coverage and omitted a substantial number of classes, whereas o1-preview achieved 100\% coverage.
    \item \textbf{Implicit Classes.} Modeling implicit business logic remains a bottleneck. Across the three projects, there are seven implicit classes in total. GPT-4 covered 28.6\%, o1-preview covered 42.9\%, and DeepSeek-R1 covered 71.4\%. In the Online Movie scenario, for example, o1-preview identified only 50\% of the implicit classes, indicating a limited ability to capture extended scenarios and implicit rules. DeepSeek-R1 performed relatively better in this regard.
    \item \textbf{Management Classes.} Most LLMs neglect management-related concerns associated with system operation and non-functional requirements. Across the three projects, there are ten management-related items in total. GPT-4 and DeepSeek-R1 achieved 0\% coverage, whereas o1-preview covered one item (10\% overall and 13\% within the corresponding project), revealing a common weakness among the evaluated models.
    \item \textbf{Impact of Readability.} In the highly readable Parking Lot scenario, which received a readability score of 81.1, o1-preview achieved the highest core-class coverage (100\%). GPT-4 and DeepSeek-R1, however, performed only moderately in the same scenario. Notably, DeepSeek-R1 also achieved 100\% core-class coverage in the less readable Online Movie scenario, which received a readability score of 73.2. These results suggest that readability affects models differently and does not exhibit a consistently positive, monotonic relationship with coverage. Further analysis of model characteristics and prompting strategies is therefore warranted.
\end{itemize}

\setcounter{table}{1}
\begin{table*}[htbp]
  \centering
  \caption{Coverage Comparison of Class Generation between Software Engineering Experts and LLMs}
  \label{tab:llm_expert_class_coverage_table2}  
  \resizebox{\linewidth}{!}{%
  \begin{tabular}{l c l c c c c}
    \toprule
    \textbf{Project Name} & \textbf{Requirement Readability} & \textbf{Class Type} & \textbf{Experts} & \textbf{GPT-4} & \textbf{o1-preview} & \textbf{DeepSeek-R1} \\
    \midrule
    \multirow{3}{*}{\makecell{Parking Lot }}& \multirow{3}{*}{81.1 (Easy to Read)} & Core Classes & 25 & 12 (48\%) & 25 (100\%) & 14 (56\%) \\
    & & Implicit Classes & 3 & 1 (33\%) & 1 (33\%) & 3 (100\%) \\
    & & Management Classes & 1 & 0 (0\%) & 0 (0\%) & 0 (0\%) \\
    \midrule
    \multirow{3}{*}{\makecell{Online Movie }}& \multirow{3}{*}{73.2 (Relatively Easy to Read)} & Core Classes & 13 & 9 (69\%) & 11 (85\%) & 13 (100\%) \\
    & & Implicit Classes & 2 & 0 (0\%) & 1 (50\%) & 1 (50\%) \\
    & & Management Classes & 8 & 0 (0\%) & 1 (13\%) & 0 (0\%) \\
    \midrule
    \multirow{3}{*}{\makecell{Stack Overflow }}& \multirow{3}{*}{75.1 (Relatively Easy to Read)} & Core Classes & 9 & 7 (78\%) & 8 (89\%) & 9 (100\%) \\
    & & Implicit Classes & 2 & 1 (50\%) & 1 (50\%) & 1 (50\%) \\
    & & Management Classes & 1 & 0 (0\%) & 0 (0\%) & 0 (0\%) \\
    \bottomrule
  \end{tabular}
  }
\end{table*}

The results demonstrate that LLMs differ substantially in their ability to identify core classes, with o1-preview exhibiting the most robust performance. In contrast, implicit and management classes remain common weaknesses across the evaluated models. Moreover, the effect of prompt readability on class-generation quality is strongly model-dependent.

\subsubsection{Relationship Correctness Analysis}\label{competency4UC}
SDMetrics employs six metrics—EC\_Attr, EC\_Par, IC\_Attr, IC\_Par, Dep\_Out, and Dep\_In—to quantify the relationships associated with each class in a class diagram~\citep{sdmetricsmanual}. We sum these values for each class to obtain a composite coupling score, with a higher score indicating stronger coupling with other classes. Classes with the highest composite scores are considered key classes because their high degree of coupling warrants particular attention during system design and maintenance.

Using the class relationships in the expert-crafted models as the reference standard, we assess the correctness of the relationships associated with the key classes that have the highest composite scores. For this purpose, we define a semantic matrix of relationship error types of class diagrams based on \citep{giannouris2026nomad}, as presented in Table \ref{tab:error_type_semantic_matrix}.

\begin{table}[htbp]
  \centering
\caption{Semantic Matrix of Relationship‑Error Types for Class Diagrams}
  \label{tab:error_type_semantic_matrix}
  \resizebox{0.9\linewidth}{!}{
  \begin{tabular}{l l l p{6cm}}
    \toprule
    \textbf{Type} & \textbf{Essence of Error} & \textbf{Architecture Impact Level} & \textbf{Description of Architecture Impact} \\
    \midrule
    t1 & Association $\rightarrow$ Composition/Aggregation & High (Level 2) & Misclassification of relationship types results in inappropriate coupling between modules \\
    t2 & Composition $\rightarrow$ Association & Critical (Level 1) & Impairs object life-cycle management and constitutes a critical error \\
    t3 & Composition $\rightarrow$ Aggregation & Medium (Level 3) & Reduces the strength of the relationship while preserving the overall structure \\
    t4 & Missing necessary association & Critical (Level 1) & Directly affects functional completeness and constitutes a critical error \\
    t5 & Redundant association & Low (Level 4) & Primarily affects system maintainability \\
    \bottomrule
  \end{tabular}
  }
\end{table}

Differentiated weights are assigned according to the severity of each error type's impact on the system architecture. To mitigate subjective bias arising from engineering experience, the initial weights were further calibrated through cross-review and consensus scoring by the three authors of this paper. All three authors have professional backgrounds in software engineering and extensive research experience in UML system modeling and software architecture design. The weight calculation formula is as follows:

The total weight score is calculated as: \[
\text{Total Weight Score} = \sum (n_i \cdot w_i)
\]

where $n_i$ denotes the number of occurrences of error type $i$, $w_i$ denotes its corresponding weight coefficient, and $i \in \{t1, t2, t3, t4, t5\}$. The weights assigned to t1, t2, t3, t4, and t5 are 0.3, 0.4, 0.2, 0.4, and 0.1, respectively, based on their impact on the architecture. A higher total weighted score indicates greater harm to the system architecture.

Table \ref{tab:relationship_error_weight} presents the weighted results for relationship errors in the class diagrams generated by the LLMs. Error types t1 and t2 account for the largest proportions, followed by t4, whereas t5 occurs only in specific scenarios and t3 exhibits model-specific characteristics. t1 (misjudged strong associations) and t2 (misjudged combinations) appear in nearly all LLM-generated class diagrams across the three projects, making them the highest-risk error types and the primary targets for remediation. t4 (missing functionality) occurs in the outputs of GPT-4 and DeepSeek-R1 and represents another high-impact error type. t5 (redundant associations) appears primarily in GPT-4's results, whereas t3 (weakened combinations) is concentrated in o1-preview's outputs for the Parking Lot project. The distribution of risk depends on both the model and the application scenario. Model performance also varies across projects: DeepSeek-R1 achieves the lowest risk score in the Parking Lot project (0.8), GPT-4 achieves the lowest score in the Online Movie project (1.4), and o1-preview achieves the lowest score in the Stack Overflow project (1.5). These results indicate that there is no ``single optimal'' model. Averaged across the three projects, DeepSeek-R1 has the lowest overall risk score ($\approx 1.4$), followed by o1-preview ($\approx 1.63$), while GPT-4 has the highest ($\approx 2.1$). This finding suggests that DeepSeek-R1 exhibits slightly greater stability across diverse application scenarios.

\begin{table*}[htbp]
  \centering
  \tiny  
  \caption{Weight Calculation and Analysis of Relationship Errors between Classes}
  \label{tab:relationship_error_weight}
  \resizebox{\linewidth}{!}{
  \begin{tabular}{l l c c}  
    \toprule
    \textbf{Project Name} & \textbf{LLM}& \textbf{Error Types and Occurrences} & \textbf{Total Weighted Score} \\
    \midrule
    \multirow{3}{*}{\makecell{Parking Lot\\ }} & GPT-4 & t1$\times$4$+$t4$\times$4 & 2.8 \\
    & o1-preview & t1$\times$2$+$t3$\times$4 & 1.4 \\
    & DeepSeek-R1 & t2$\times$2 & 0.8 \\
    \midrule
    \multirow{3}{*}{\makecell{Online Movie\\}} & GPT-4 & t2$\times$3$+$t5$\times$2 & 1.4 \\
    & o1-preview & t1$\times$4$+$t2$\times$2 & 2.0 \\
    & DeepSeek-R1 & t1$+$t2$\times$2$+$t4 & 1.5 \\
    \midrule
    \multirow{3}{*}{\makecell{Stack Overflow\\ }} & GPT-4 & t1$\times$4$+$t4$\times$2$+$t5 & 2.1 \\
    & o1-preview & t1$+$t2$\times$3 & 1.5 \\
    & DeepSeek-R1 & t1$+$t2$\times$4 & 1.9 \\
    \bottomrule
  \end{tabular}
  }
\end{table*}

In summary, the relationship-error weights reported in Table \ref{tab:relationship_error_weight} provide an accurate representation of the risks associated with design flaws in LLM-generated class diagrams. o1-preview maintains a relatively low risk level in most scenarios but tends to overuse combination relationships in moderately complex tasks, such as the Online Movie project. DeepSeek-R1 demonstrates robust overall performance but exhibits concentrated high-risk errors, whereas GPT-4 shows the combined effects of high-risk omissions and misjudgments in certain scenarios. These findings indicate that all LLM-generated class diagrams require repair strategies targeting the three high-risk error types: t1, t2, and t4.

\subsubsection{Inheritance Hierarchy Analysis}\label{competency4UC}
Analyzing the DIT and CLD parameters provided by SDMetrics effectively measures the complexity of class hierarchies~\citep{sdmetricsmanual}, thereby providing an important reference for system design, optimization, and maintenance.

Table~\ref{tab:inheritance_relationship_params} presents the SDMetrics statistics for inheritance-related parameters in expert-crafted and LLM-generated class diagrams. Expert-crafted class diagrams exhibit average DIT values of no more than 0.78 and average CLD values of no more than 0.33 across all three projects, with maximum DIT and CLD values of 1. These results reflect the design principle of ``shallow inheritance and controlled reuse,'' which emphasizes composition over inheritance and clearly defined responsibilities. In the business-oriented scenarios of the Parking Lot and Online Movie projects, LLMs generally produce insufficient or absent inheritance relationships. GPT-4 fails to model inheritance in either project, while o1-preview and DeepSeek-R1 exhibit substantially lower inheritance coverage than the expert-crafted diagrams. Such deficiencies may lead to code duplication, unclear hierarchies, difficulties in system extension, and the loss of polymorphic behavior. In the knowledge-oriented Stack Overflow scenario, however, some LLMs exhibit inheritance ``inflation'': GPT-4 and DeepSeek-R1 both reach maximum DIT and CLD values of 2. This pattern may indicate that their inference and generation strategies favor identifying or constructing more complex inheritance structures, thereby increasing maintenance and evolution costs. Across the three projects, o1-preview generally more closely matches expert judgments regarding inheritance depth, although differences remain in the quantity and distribution of inheritance relationships. Further refinement is therefore necessary~\citep{giannouris2026nomad,sdmetricsmanual}.

\begin{table*}[htbp]
  \centering
  \caption{SDMetrics Statistics for Inheritance-Related Parameters in Class Diagrams}
  \label{tab:inheritance_relationship_params}
  \resizebox{\linewidth}{!}{%
  \begin{tabular}{l l c c c c c}
    \toprule
    \textbf{Project Name} & \textbf{Expert/LLM} & \textbf{Number of Inherited Classes} & \textbf{Average DIT} & \textbf{Average CLD} & \textbf{Maximum DIT} & \textbf{Maximum CLD} \\
    \midrule
    \multirow{4}{*}{\makecell{Parking Lot\\ }} & Expert & 18 & 0.78 & 0.22 & 1 & 1 \\
    & GPT-4 & 0 & 0 & 0 & 0 & 0 \\
    & o1-preview & 15 & 0.80 & 0.20 & 1 & 1 \\
    & DeepSeek-R1 & 5 & 0.60 & 0.40 & 1 & 1 \\
    \midrule
    \multirow{4}{*}{\makecell{Online Movie\\}} & Expert & 12 & 0.67 & 0.33 & 1 & 1 \\
    & GPT-4 & 0 & 0 & 0 & 0 & 0 \\
    & o1-preview & 2 & 0.50 & 0.50 & 1 & 1 \\
    & DeepSeek-R1 & 3 & 0.67 & 0.33 & 1 & 1 \\
    \midrule
    \multirow{4}{*}{\makecell{Stack Overflow\\ }} & Expert & 3 & 0.67 & 0.33 & 1 & 1 \\
    & GPT-4 & 7 & 0.71 & 0.57 & 2 & 2 \\
    & o1-preview & 3 & 0.67 & 0.33 & 1 & 1 \\
    & DeepSeek-R1 & 4 & 0.75 & 1.00 & 2 & 2 \\
    \bottomrule
  \end{tabular}
  }
\end{table*}

In summary, LLMs exhibit inconsistent performance in identifying inheritance relationships during automated class diagram generation. Some models, such as GPT-4, may omit inheritance relationships entirely, whereas others, such as DeepSeek-R1, may introduce unwarranted multilevel inheritance structures. Achieving more robust results requires incorporating contextual information from the input text, leveraging relevant domain knowledge, and systematically validating and refining the generated outputs.

\subsubsection{Design Rule Compliance Analysis}\label{Design Rule Compliance}
SDMetrics~\citep{wust2005sdmetrics} performs rule-based conformance checks on class diagrams using its built-in detector, which evaluates designs against established software engineering principles. Table~\ref{tab:sdmetrics_class_rule_errors} summarizes the typical error types identified during these inspections. The design rationale underlying the selected rules is as follows. The \textit{Unused} rule promotes concise class diagram designs by identifying redundant elements that may otherwise result in dead code and unnecessarily complicate the system architecture. The \textit{DepCycle} rule supports modular independence by detecting dependency cycles, thereby reducing intermodule coupling and improving the overall maintainability of the system. The \textit{AttrNameOvr} rule detects attribute-naming conflicts between parent and child classes, reducing ambiguity during code comprehension and improving code readability and maintainability~\citep{sdmetricsmanual}.

\begin{table*}[htbp]
  \centering
  \footnotesize  
  \caption{Common Errors Identified by SDMetrics in Class Rule Checking}
  \label{tab:sdmetrics_class_rule_errors}
  \resizebox{\linewidth}{!}{
  \begin{tabular}{l l l p{7cm}}  
    \toprule
    \textbf{Violation Name} & \textbf{Category} & \textbf{Severity} & \textbf{Description} \\
    \midrule
    Unused (Unused Class) & Completeness & 1-High & The class is not used anywhere \\
    DepCycle (Cyclic Dependency) & Style & 2-Med & The class has cyclic references \\
    AttrNameOvr (Attribute Override) & Naming & 2-Med & The class defines an attribute with the same name as an inherited attribute \\
    \bottomrule
  \end{tabular}
  }
\end{table*}

Table~\ref{tab:class_diagram_rule_detection} presents the detected design-rule violations in the class diagrams generated by experts and LLMs. To account for differences in the severity of violation types, we compute a severity-weighted total score to quantitatively assess the architectural risks associated with these violations. The severity weights are defined as follows: high-severity violations receive 3 points (e.g., \textit{Unused}), whereas medium-severity violations receive 2 points (e.g., \textit{DepCycle} and \textit{AttrNameOvr}).

The results in Table~\ref{tab:class_diagram_rule_detection} show that the expert-crafted diagrams and those produced by DeepSeek-R1 achieve the best performance, with no detected violations and weighted total scores of zero. Both approaches are therefore suitable for engineering scenarios with stringent requirements for design quality. By contrast, GPT-4, the primary optimization model, exhibits multiple violations across the evaluated projects, including \textit{Unused}, \textit{DepCycle}, and \textit{AttrNameOvr}, and records the highest weighted total score (13). These results indicate limitations in its ability to enforce structural design constraints. In particular, cyclic dependencies should be addressed through design strategies such as dependency injection and interface-based isolation, while the generation of redundant classes should also be reduced. o1-preview would benefit from improved contextual reasoning and association capabilities to prevent the generation of unused classes at the source.

\begin{table*}[htbp]
  \centering
  \normalsize  
  \caption{Design‑Rule Violation Detection Results for Expert‑Crafted and LLM‑Generated Class Diagrams}
  \label{tab:class_diagram_rule_detection}
  \resizebox{\linewidth}{!}{
  \begin{tabular}{l l l l c c}
    \toprule
    \textbf{Model (Expert/LLM)} & \textbf{Parking Lot} & \textbf{Online Movie} & \textbf{Stack Overflow} & \textbf{Total Violations} & \textbf{Severity-Weighted Score} \\
    \midrule
    Expert & \makecell{Unused: 0\\DepCycle: 0} & \makecell{Unused: 0\\DepCycle: 0} & \makecell{Unused: 0\\DepCycle: 0} & 0 & 0 \\
    GPT-4 & \makecell{Unused: 1 (3)\\DepCycle: 0} & \makecell{Unused: 0\\DepCycle: 2 (4)} & \makecell{Unused: 0\\DepCycle: 0 (0)\\AttrNameOvr: 3 (6)} & 6 & 13 \\
    o1-preview & \makecell{Unused: 1 (3)\\DepCycle: 0} & \makecell{Unused: 0\\DepCycle: 0} & \makecell{Unused: 0\\DepCycle: 0} & 1 & 3 \\
    DeepSeek-R1 & \makecell{Unused: 0\\DepCycle: 0} & \makecell{Unused: 0\\DepCycle: 0} & \makecell{Unused: 0\\DepCycle: 0} & 1 & 3 \\
    \bottomrule
  \end{tabular}
  }
\end{table*}

In summary, although SDMetrics identifies issues such as unused classes and cyclic dependencies in class diagrams, it does not provide remediation strategies. This paper leverages the results of automated detection to guide the targeted repair of design-rule violations, thereby improving the standardization and quality of LLM-generated class diagrams.

\begin{mdframed}
    \textbf{Key Findings of RQ1}: We found that LLM-generated class diagrams generally underperform expert-crafted diagrams in terms of modeling quality. Substantial differences exist among models in identifying core classes, while the modeling of implicit and administrative classes remains a common weakness across all evaluated LLMs. The effect of prompt readability on diagram-generation quality also varies considerably across models. All LLMs exhibit specific high-risk errors in relationship modeling, requiring targeted repairs for three critical violation types: t1, t2, and t4 (see Table~\ref{tab:error_type_semantic_matrix}). Inheritance modeling accuracy varies widely, ranging from frequent omissions (e.g., GPT-4) to excessive expansion (e.g., DeepSeek-R1). In addition, some LLMs generate design-rule violations, such as unused classes and cyclic dependencies, highlighting the need for targeted remediation and further optimization.
\end{mdframed}

\subsection{Answer to RQ2}\label{RQ2}
Through a comparative study of expert-crafted and LLM-generated UML class diagrams, we identify four recurring defects: inadequate recognition of key classes, high-risk relationship errors, missing or excessive inheritance, and violations of design rules. To answer RQ2, we develop problem-specific repair strategies based on domain-knowledge injection, memory reinforcement, and rule-based targeted repair. We then empirically evaluate the effectiveness of these strategies in mitigating common defects in LLM-generated class diagrams.

\subsubsection{Core Class Repair}\label{RQ2:competency4UA}
\textbf{Repair Method}.To address the core-class modeling defects summarized in RQ1—including missing essential classes and misidentified entity types in vanilla LLM-generated UML diagrams—we employ a prompt-driven external knowledge-injection procedure for systematic diagram rectification. This procedure adopts fixed prompt configurations and comprises three staged knowledge-guided operations. First, multi-type entity categorization ($P_{\mathrm{Cat}}$) defines five fundamental modeling entity categories, establishing standardized domain classification rules to distinguish physical entities, role entities, events, interactions, and conceptual entities. Second, requirement-driven core class identification ($P_{\mathrm{Ident}}$) applies the formulated classification criteria to extract and validate all system-required entity classes from natural language requirements, ensuring complete and accurate core entity screening. Third, standard diagram generation and rectification ($P_{\mathrm{Gen}}$) reconstructs UML class diagrams with the validated core classes and outputs standardized PlantUML code, repairing defects of core class omission and misidentification. Expert core-class annotations from three benchmark projects (Parking Lot, Online Movie, and Stack Overflow) serve as ground-truth references. Consistent prompt settings are applied across GPT-4, o1-preview, and DeepSeek-R1 to ensure experimental reproducibility and fair model evaluation.

\textbf{Evaluation Method}. We conduct a double-blind expert evaluation with inter-rater consistency assessed using Fleiss' $\kappa$. The three authors independently evaluate the repaired class diagrams without knowledge of the model identities or generation order. The diagrams are scored along three dimensions. First, \emph{Semantic Consistency of Class Names} is rated as follows: 0 = inconsistent or misleading, 1 = approximately consistent or synonymous but imprecise, and 2 = fully consistent. Second, \emph{Attribute/Type Matching} comprises two components—attribute coverage and type accuracy—each scored on a 0--2 size; the two scores are then averaged and rounded to the nearest integer. Third, \emph{Design Principle Compliance}, evaluated with respect to the Single Responsibility Principle (SRP), is scored as follows: 0 = non-compliant, 1 = generally compliant with minor overreach, and 2 = explicitly compliant. For each evaluation, the reviewers provide evidence linking the judgment to the requirements text, as well as evidence from the relevant class attributes and methods. Fleiss' $\kappa$ is then calculated to assess inter-rater consistency and evaluation reliability~\citep{solid2025empirical}.

\textbf{Core Class Achievement Rate}. This indicator measures the degree to which system core classes are correctly identified and repaired, using the expert-crafted core classes as the reference standard. It is calculated as the proportion of fully compliant repaired core classes among all target core classes:

\begin{equation*}
R_{class} = \frac{N_{corrected\_class}}{N_{target\_class}} \times 100\%
\end{equation*}

where $N_{\mathrm{corrected\_class}}$ denotes the number of core classes that fully satisfy the expert benchmark criteria after repair, including semantic consistency of the class name, attribute--type matching, and compliance with the Single Responsibility Principle (SRP). $N_{\mathrm{target\_class}}$ denotes the total number of expert-crafted target core classes in each test case. This metric is also referred to as \emph{core-class coverage} in the subsequent analysis.

\textbf{Determination Criteria}. A repaired core class is considered \emph{Successfully Fixed} when, based on the average scores assigned by the three reviewers, all three dimensions meet or exceed the minimum threshold (i.e., class-name consistency, attribute--type matching, and SRP compliance are each $\ge 1$), the total score is at least $5/6$, and inter-rater consistency satisfies $\kappa \ge 0.60$. If any dimension scores below 1 or $\kappa < 0.60$, the class is initially deemed \emph{Failed to Fix} and undergoes a secondary review involving Delphi-style discussion and evidence verification before a final determination is made.

\begin{figure}[!t]
\centering
\includegraphics[width=1\textwidth]{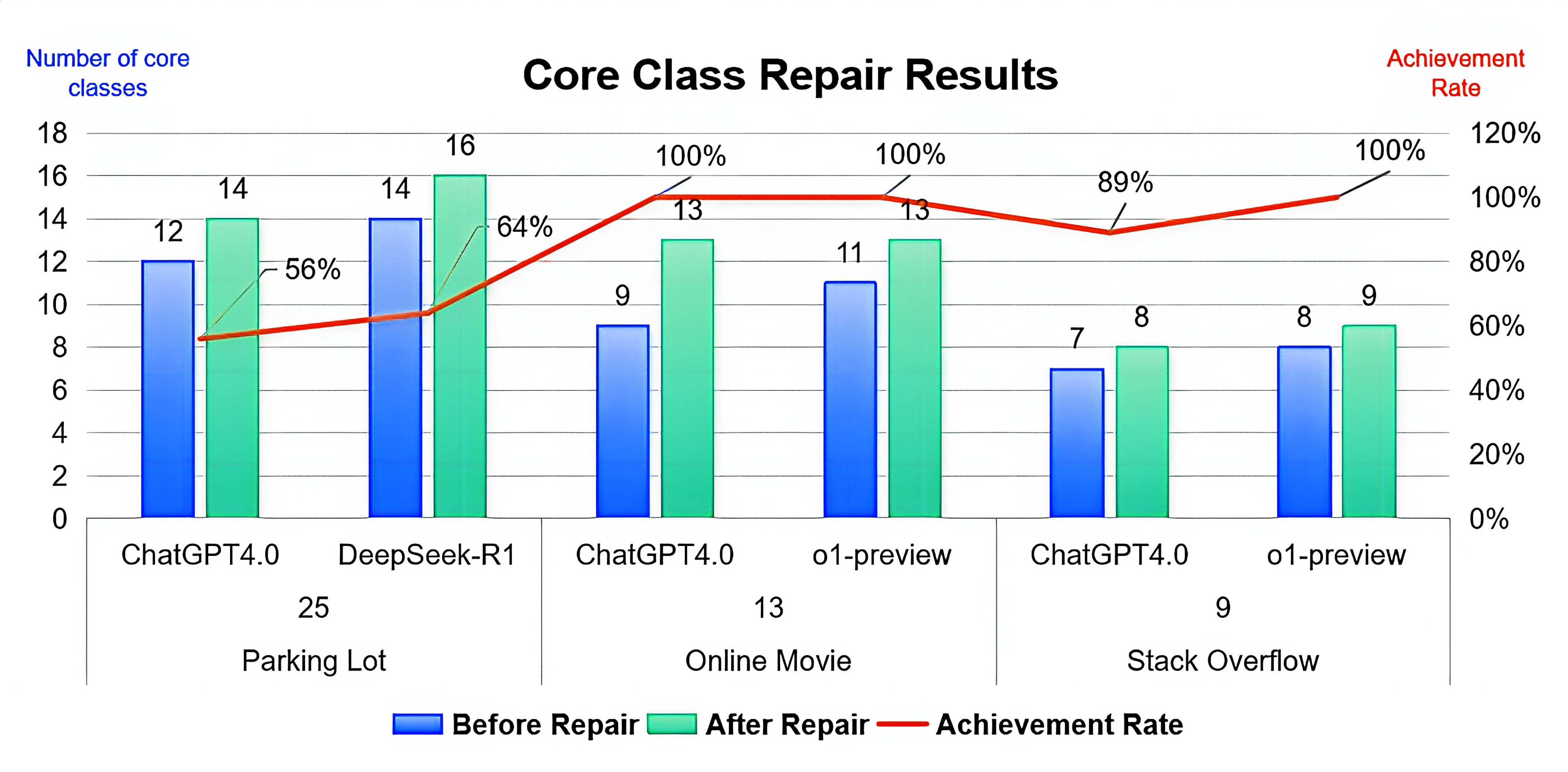}
\caption{Core Class Repair Results}
\label{fig:scorediff1}
\end{figure}

\textbf{Quantitative Results of Core Class Repair}. Figure~\ref{fig:scorediff1} summarizes the core-class repair outcomes across the three case systems. Compared with the original automatically generated class diagrams, the repaired models consistently achieve higher core-class coverage. We quantify this improvement using the achievement rate (i.e., core-class coverage), denoted by $R_{\mathrm{class}}$, which is defined as the number of core classes identified after repair divided by the total number of target core classes. For the Parking Lot system, which contains 25 target core classes, GPT-4 and DeepSeek-R1 each identified two additional core classes, increasing their counts from 12 to 14 and from 14 to 16, respectively. Their post-repair coverage rates were therefore 56\% and 64\%, although neither achieved full coverage. For the Online Movie system, which contains 13 target core classes, both GPT-4 and o1-preview achieved complete coverage after repair (13/13), representing improvements of 4 and 2 classes, respectively. For Stack Overflow, which contains 9 target core classes, GPT-4 increased its coverage from 7 to 8 classes (89\%), while o1-preview increased its coverage from 8 to 9 classes (100\%); both improvements correspond to one additional core class. Averaged as an unweighted mean across case--model pairs, core-class coverage increased from approximately 71\% before repair to approximately 85\% after repair, underscoring the effectiveness of the proposed repair procedure. 

\subsubsection{Coupling Relationship Repair}\label{competency4UC}
\textbf{Repair Method}. To address ambiguous, misconfigured, and inaccurate class coupling relationships in initial LLM-generated UML diagrams summarized in RQ1, this study adopts a memory-reinforced repair paradigm to optimize relationship modeling, since conventional external knowledge injection yields limited improvement on complex coupling defects. This paradigm adopts fixed multi-stage prompt configurations and retrieves implicit domain knowledge from LLM parameters to form a valid knowledge memory set for iterative diagram rectification. It comprises six logically progressive stages. First, relationship definition clarification ($P_{\mathrm{Def}}$) clarifies the fundamental definitions of typical UML class relationships, including dependency, association, aggregation, and composition, establishing basic conceptual boundaries. Second, graphical representation differentiation ($P_{\mathrm{Repr}}$) distinguishes the graphical features of different relationships in class diagrams, covering connection styles and arrow conventions. Third, semantic scenario discrimination ($P_{\mathrm{Sem}}$) analyzes the practical semantic differences of various coupling relationships in real application contexts. Fourth, multi-dimensional knowledge review ($P_{\mathrm{Rev}}$) integrates and summarizes the differentiated rules of UML relationships from definition, representation, and semantic perspectives. Fifth, diagram-oriented structural analysis ($P_{\mathrm{Anal}}$) examines the functional responsibilities of individual classes and their existing interaction patterns based on the unrepaired PlantUML class diagram. Sixth, targeted relationship redesign ($P_{\mathrm{Redes}}$) rearranges class connection modes and corrects inappropriate graphical features according to summarized discrimination rules and diagram analysis results, so as to precisely restore the actual coupling relationships between classes. Equivalent prompt settings are deployed across GPT-4, o1-preview, and DeepSeek-R1 to ensure stable reasoning outputs, iterative repair validity, and consistent experimental standards.

\textbf{Evaluation Method}. Using expert-crafted class diagrams as the gold standard, we assess repair effectiveness through a two-tier framework that combines prioritization of critical classes with error weighting. In Tier 1, we adopt SDMetrics to compute the comprehensive coupling index \( CI(c) = EC_{\text{Attr}} + EC_{\text{Par}} + IC_{\text{Attr}} + IC_{\text{Par}} + Dep_{\text{Out}} + Dep_{\text{In}} \) for each class.We prioritize the critical class set with the highest $CI$ values, thereby focusing verification on the relationships with the greatest potential impact on the system architecture. In Tier 2, we annotate the differences between the repaired class diagram and the gold standard using an error-type semantic matrix: $t_1$ (Association $\rightarrow$ Composition/Aggregation), $t_2$ (Composition $\rightarrow$ Association), $t_3$ (Composition $\rightarrow$ Aggregation), $t_4$ (Missing Required Association), and $t_5$ (Redundant Association).We apply the weight vector \(w = \{0.3, 0.4, 0.2, 0.4, 0.1\}\). The project--model‑level weighted error score is defined as \(E = \sum_i n_i w_i\), where \(n_i\) denotes the occurrence count of error type \(i\). A lower value of \(E\) indicates that the repaired relationships more closely conform to the gold standard. Ambiguous matches undergo dual-review verification to ensure alignment and scoring consistency~\citep{llmrepair2025survey}.

\textbf{Coupling Relationship Repair Rate}. This metric quantifies the overall remediation of relationship-semantic errors from a defect-reduction perspective. It uses the relative reduction in the weighted architectural error score to measure the improvement in the rationality of system coupling:

\begin{equation*}
R_{coupling} = \frac{E_{before} - E_{after}}{E_{before}} \times 100\%
\end{equation*}

where \(E_{\mathrm{before}}\) and \(E_{\mathrm{after}}\) denote the total weighted error scores before and after repair, respectively. This indicator is also referred to as the \emph{repair rate} in the subsequent analysis.

\textbf{Determination Criteria}. A substantial improvement is considered to have been achieved when either \(R_{\mathrm{coupling}} \ge 50\%\) or the number of relationship errors in the critical categories is reduced by at least half. If only one of these criteria is satisfied, the result is classified as \emph{Partially Fixed}; if neither criterion is satisfied, it is classified as \emph{Unfixed}. 

\begin{figure}[!t]
\centering
\includegraphics[width=1\textwidth]{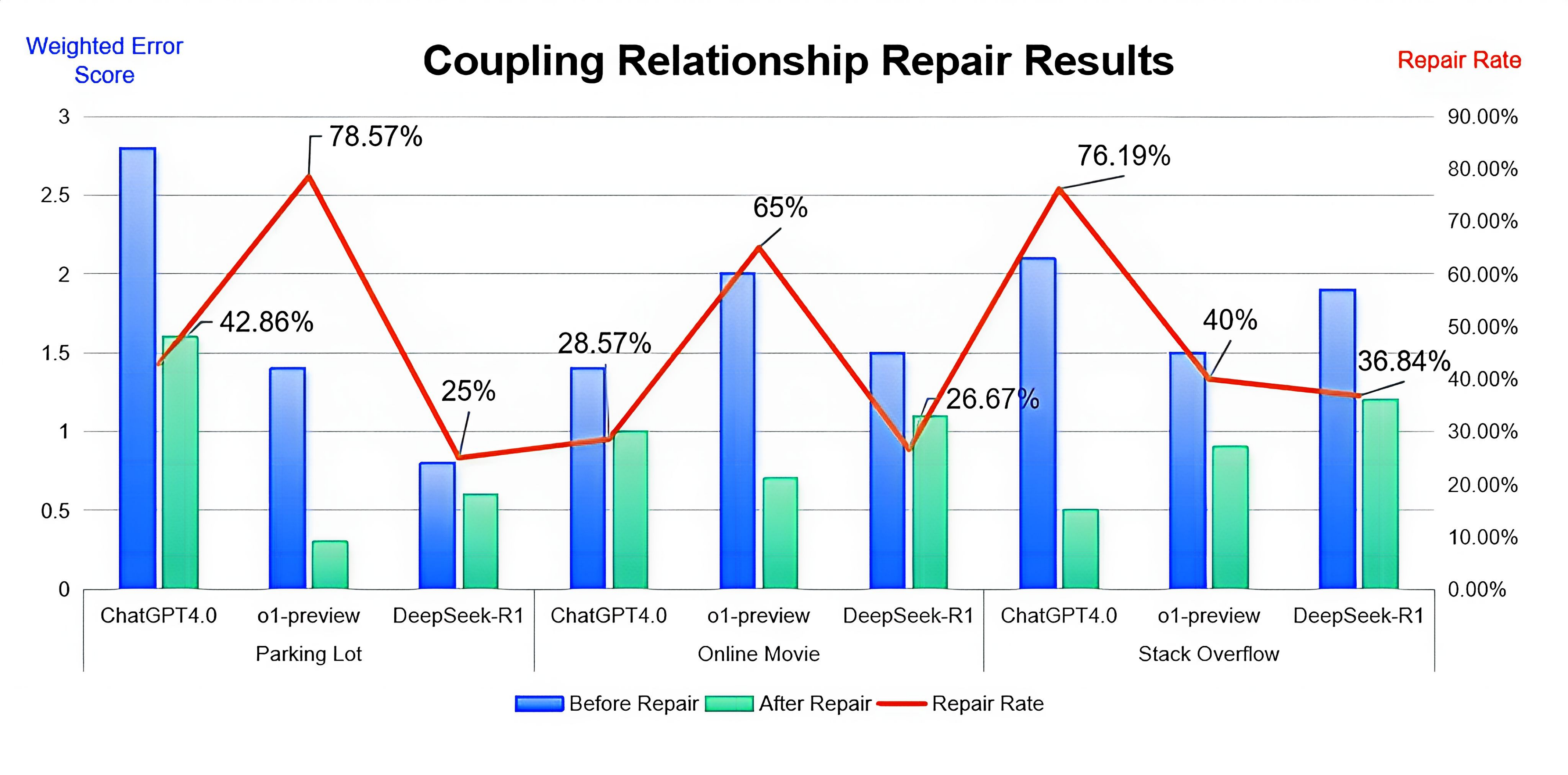}
\caption{Coupling Relationship Repair Results}
\label{fig:scorediff2}
\end{figure}

\textbf{Quantitative Results of Coupling Relationship Repair}. Figure~\ref{fig:scorediff2} reports the relationship-repair results for the LLM-generated class diagrams. Using the memory-reinforcement procedure, o1-preview reduced the weighted error score, $E$, from 1.4 to 0.3 in the Parking Lot case, corresponding to a repair rate of 78.57\% (i.e., $R_{\mathrm{coupling}}$, the proportional reduction in $E$ from before to after repair). In the Stack Overflow case, GPT-4 reduced $E$ from 2.1 to 0.5, yielding a repair rate of $R_{\mathrm{coupling}}=76.19\%$ and likewise indicating substantial improvement. By contrast, DeepSeek-R1 exhibited the smallest gains among the evaluated cases. In the Online Movie case, its $E$ score decreased from 1.5 to 1.1, corresponding to a repair rate of $R_{\mathrm{coupling}}=26.67\%$; its post-repair score also remained higher than that of GPT-4. The principal failure mode involved downgrading high-severity errors to medium-severity errors (e.g., $t_2\rightarrow t_3$) rather than eliminating them. Overall, these results demonstrate the benefits of memory reinforcement while indicating that all models still have room for improvement in relationship repair. Across the three evaluated cases, the coupling relationship repair rate was 46\%, reflecting the overall average performance of the LLMs in repairing structural coupling defects.

A post-hoc analysis indicates that DeepSeek-R1 underperforms in scenarios requiring precise discrimination of relationship semantics. In the Parking Lot case, the expected composition between ``ParkingLot'' and ``Display'' was misclassified as an association and was upgraded only to an aggregation after repair. This result suggests insufficient recognition of strong ownership and lifecycle binding, as well as the absence of explicit prompt rules—for example, rules specifying that physical component relationships should be modeled as compositions. In the Online Movie case, the model conflated the strong dependence between ``Movie'' and ``Screening'' (composition) with the weak association between ``User'' and ``Order''. This mismatch persisted after repair, suggesting difficulty in handling coexisting strong and weak ties among multiple entities, possibly because of limited exposure to analogous business patterns. In Stack Overflow, the compositions between ``Question--Comment'' and ``Question--Answer'' were treated as associations and remained uncorrected. This outcome indicates limited understanding of the tight dependence between a content entity and its attachments, as well as insufficient pattern-based guidance in the prompts—for example, guidance that content aggregation implies composition. Together, these factors hindered effective correction.

Overall, model repair led to varying degrees of improvement in coupling-relationship correctness and was accompanied by a substantial reduction in weighted architectural error scores. The evaluations above focus on objective error severity and the effects of structural-defect remediation, based on cross-calibrated error-type weights. Comprehensive multidimensional architectural-quality quantification based on customized quality coefficients will be further developed and validated in future research~\citep{locobench2025benchmark}.

\subsubsection{Inheritance Relationship Repair }\label{competency4UC}
\textbf{Repair Method}. To address the two principal issues identified by the structural-complexity analysis in RQ1—missing or sparse inheritance relationships and excessively deep inheritance hierarchies (DIT $\geq 2$ or CLD $\geq 2$)—we employ a memory-enhanced procedure to systematically restructure generalization and realization relationships in UML class diagrams. The procedure uses a unified, fixed prompt configuration and comprises three stages. First, \emph{memory activation and value alignment} ($P_{\mathrm{InhDef}}$) revisits the definition of inheritance and its potential benefits, including abstraction and reuse, polymorphism, and the enforcement of consistency constraints. This stage produces an actionable set of rules for determining when and how inheritance should be used. Second, \emph{risk framing and metric binding} ($P_{\mathrm{Metr}}$) clarifies the meanings of DIT and CLD, explains how DIT $> 2$ or CLD $> 2$ may increase complexity, fragility, and maintenance costs, and establishes control objectives and warning thresholds. These objectives include eliminating over-threshold inheritance chains, reducing hierarchy depth, and avoiding the misuse of inheritance. Third, \emph{diagnosis and guided refactoring} ($P_{\mathrm{Redes}}$) examines the unrepaired PlantUML diagram to identify missing inheritance relationships, inheritance used solely for code reuse, and excessively deep chains that cause DIT or CLD to exceed the specified thresholds. The rules established in the first two stages are then applied to refactor the diagram by extracting or consolidating abstract parent classes, introducing interfaces to separate variation points, replacing misused inheritance with composition, and flattening overly deep hierarchies. All prompt settings are held constant across the evaluated models, including GPT-4, o1-preview, and DeepSeek-R1, to ensure reproducibility.

\textbf{Evaluation Method}. We use the expert-crafted class diagram as the gold standard and evaluate the repaired diagrams along two dimensions: correctness and structural complexity. Correctness is assessed through binary matching against the expert-crafted generalization and realization relationships. After synonym normalization and node mapping, each repaired inheritance edge is checked for exact agreement with the gold standard in terms of both the child--parent pair and the relationship type. Matching edges are labeled \emph{correct}, whereas non-matching edges are labeled \emph{incorrect}. Structural complexity is assessed using SDMetrics by comparing the pre- and post-repair distributions of DIT and CLD. The reported statistics include the mean, median, maximum, quartiles, and proportion of classes exceeding the specified thresholds. Particular emphasis is placed on changes in the proportion of classes with DIT $> 2$ or CLD $> 2$, thereby quantifying the extent to which excessive hierarchy depth is reduced.

\textbf{Inheritance Group Achievement Rate}. This indicator measures the structural completeness of the repaired inheritance hierarchies. It is defined as the proportion of correctly repaired inheritance groups that conform to the expert-crafted benchmarks:

\begin{equation*}
R_{inherit} = \frac{N_{corrected\_inherit}}{N_{target\_inherit}} \times 100\%
\end{equation*}

where $N_{\mathrm{corrected\_inherit}}$ denotes the number of inheritance groups that fully match the expert-crafted standards after repair, and $N_{\mathrm{target\_inherit}}$ denotes the total number of expert-crafted target inheritance groups in each test case. This metric is referred to as the \emph{achievement rate} in the subsequent analysis.

\textbf{Determination Criteria}. The determination is based solely on binary matching against the expert-crafted standard. At the relation level, a target inheritance edge is considered \emph{Successfully Repaired} if it matches the expert-crafted class diagram in both the class pair and the relationship type. At the diagram level, a repair is considered \emph{Successful} only if all target inheritance relationships match the expert-crafted class diagram, corresponding to $R_{\mathrm{inherit}}=100\%$. Any non-matching relationship results in a verdict of \emph{Not Repaired}. When naming ambiguities arise, the evaluation must include documented normalization and node mapping, together with an evidence chain based on class responsibilities and relationship semantics. Evaluators review this evidence to ensure traceability and consistency. Complexity statistics, including DIT and CLD, are reported as supplementary evidence of structural improvement but are not used to determine repair success.

\begin{figure}[!t]
\centering
\includegraphics[width=1\textwidth]{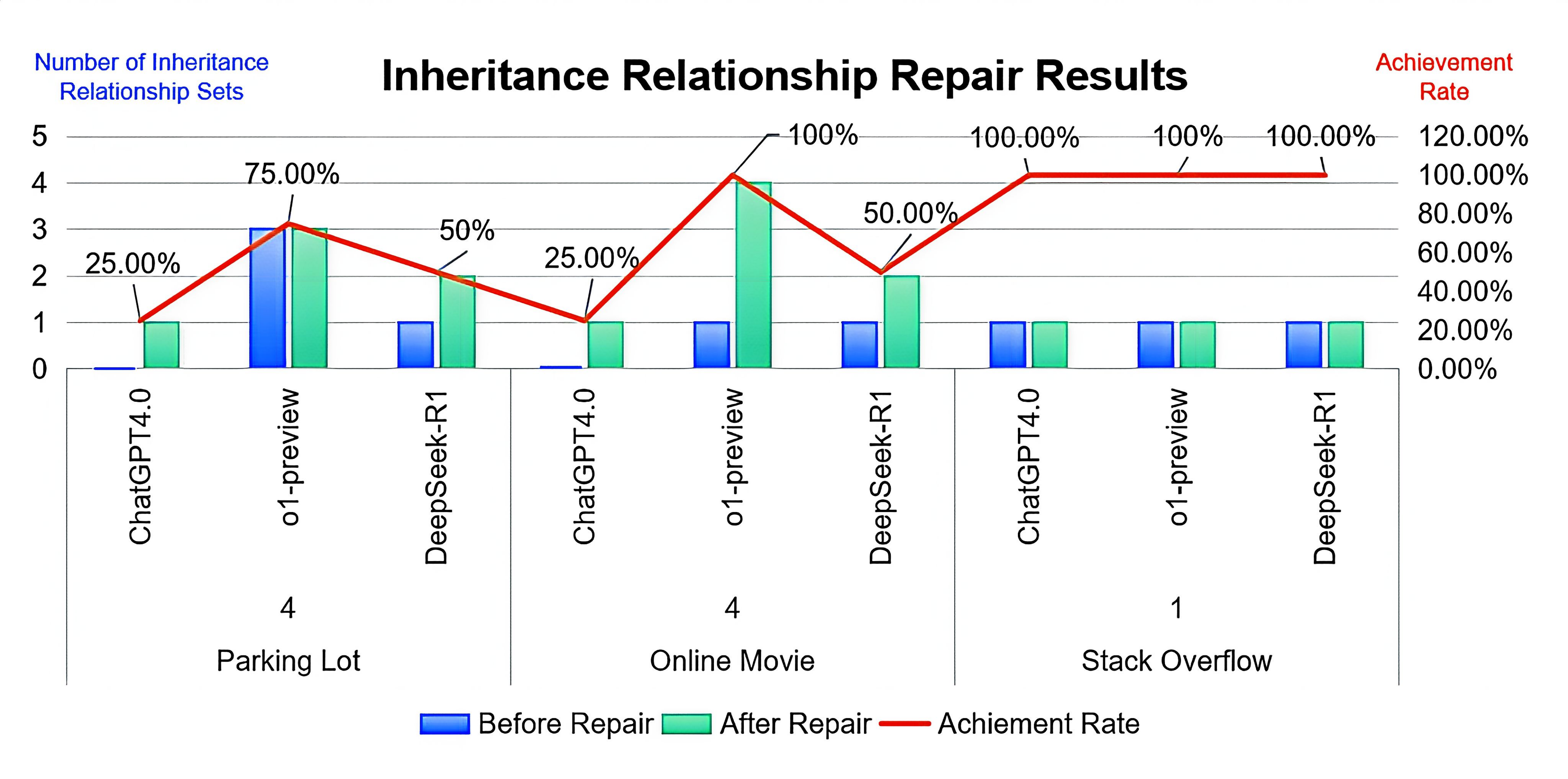}
\caption{Inheritance Relationship Repair Results}
\label{fig:scorediff3}
\end{figure}

\textbf{Quantitative Results of Inheritance Relationship Repair}. Figure~\ref{fig:scorediff3} presents the inheritance-repair results for the LLM-generated UML class diagrams. We use the inheritance group achievement rate, denoted by $R_{\mathrm{inherit}}$, to quantify the proportion of expert-crafted inheritance groups correctly reproduced by each model after repair.

In the Parking Lot case, the gold standard contained four target inheritance groups ($N_{\mathrm{target\_inherit}}=4$). GPT-4 identified none before repair and one after repair, yielding an achievement rate of $R_{\mathrm{inherit}}=25\%$. o1-preview identified three groups both before and after repair, corresponding to $R_{\mathrm{inherit}}=75\%$. DeepSeek-R1 improved from one group before repair to two groups after repair, achieving $R_{\mathrm{inherit}}=50\%$.

The Online Movie case likewise contained four target inheritance groups ($N_{\mathrm{target\_inherit}}=4$). GPT-4 identified one group both before and after repair, resulting in $R_{\mathrm{inherit}}=25\%$. o1-preview improved from one group before repair to all four groups after repair, thereby achieving $R_{\mathrm{inherit}}=100\%$. DeepSeek-R1 improved from one group to two groups, yielding $R_{\mathrm{inherit}}=50\%$.

In the Stack Overflow case, the gold standard contained one target inheritance group ($N_{\mathrm{target\_inherit}}=1$). All three models—GPT-4, o1-preview, and DeepSeek-R1—identified this group both before and after repair, resulting in $R_{\mathrm{inherit}}=100\%$ for each model. Overall, o1-preview outperformed the other models in reproducing the target inheritance structures in the Parking Lot and Online Movie cases. Across all evaluated model--project combinations, the average inheritance group achievement rate was approximately 69\%.

Although the number of inheritance groups generated by GPT-4 and DeepSeek-R1 remained unchanged in the Stack Overflow case, classes with DIT $\geq 2$ and CLD $\geq 2$ were observed before repair. After repair, the maximum values of both DIT and CLD decreased to 1, indicating a reduction in inheritance-hierarchy depth and structural complexity. Table~\ref{tab:inheritance_repair_avg} summarizes the cross-project inheritance-repair results averaged across the evaluated models. 

\begin{table}[htbp]
  \centering
  \caption{Inheritance Repair Results Averaged across Multiple Projects for Each Evaluated LLM}
  \label{tab:inheritance_repair_avg}
  \begin{tabular}{>{\centering\arraybackslash}p{2.3cm} >{\centering\arraybackslash}p{3.0cm} >{\centering\arraybackslash}p{1.8cm} >{\centering\arraybackslash}p{1.8cm} >{\centering\arraybackslash}p{1.2cm}}
    \hline
    LLM         & Average Increase in Inheritance Sets & DIT Reduction Rate & CLD Reduction Rate & Repair Rate \\
    \hline
    GPT-4       & 0.67                                 & 16.67\%            & 16.67\%            & 50\%        \\
    o1-preview  & 1                                    & 0\%                & 0\%                & 91.67\%     \\
    DeepSeek-R1 & 0.67                                 & 16.67\%            & 16.67\%            & 66.67\%     \\
    \hline
  \end{tabular}
\end{table}

The post-repair effects on the evaluated systems can be summarized as follows. In the Online Movie case, o1-preview introduced four inheritance groups, supporting strategy-like designs for concerns such as payment methods and seat states. This structure can improve extensibility and facilitate the implementation of new features. In the Parking Lot case, DeepSeek-R1 introduced two inheritance groups that provide basic polymorphism for entities such as parking-space types, thereby improving code reuse and reducing duplication. In the Stack Overflow case, the DIT and CLD values produced by GPT-4 and DeepSeek-R1 decreased from 2 to 1, indicating shallower inheritance hierarchies and potentially improving the isolation and localization of faults during maintenance.

However, systems that did not fully satisfy the target inheritance structures may still retain design deficiencies. For example, the Parking Lot system contained only one or two inheritance groups after repair, which may leave some payment-related behaviors duplicated and consequently increase technical debt and maintenance costs. These observations suggest that inheritance repair can improve extensibility, reuse, and structural simplicity, but its benefits depend on the completeness and semantic correctness of the repaired relationships.

\subsubsection{Design Rule Violation Repair}\label{competency4UC}
\textbf{Repair Method}. To resolve design‑rule violations uncovered in RQ1, such as unused classes, cyclic dependencies, and attribute‑name overriding within LLM‑generated UML class diagrams, we adopt an automated‑driven targeted‑repair procedure. This procedure uses fixed prompt configurations and follows a two‑stage workflow. First, violation localization obtains violation information via SDMetrics, which scans the unrepaired PlantUML diagrams to detect design‑rule defects and record their concrete types and corresponding faulty elements. Second, violation‑oriented remediation ($P_{\mathrm{TgtRep}}$) takes the raw unrepaired diagram together with the detected violation description as input, and generates refactoring operations to eliminate the targeted design‑rule defect. All prompt settings are held constant across the evaluated models, including GPT‑4, o1‑preview, and DeepSeek‑R1, to ensure reproducibility.

\textbf{Evaluation Method}. Design-rule violation repair is evaluated using SDMetrics. First, we detect violations in the unrepaired diagram. We then apply the targeted prompt-based repair and re-analyze the repaired diagram using the same SDMetrics configuration. A repair is considered successful if a violation reported before repair is no longer detected after repair.

\textbf{Rule Violation Repair Rate}. This indicator measures the extent to which UML design-rule violations are eliminated. It is calculated as the proportion of originally detected violations that are successfully fixed:

\begin{equation*}
R_{rule} = \frac{N_{fixed\_rule}}{N_{original\_rule}} \times 100\%
\end{equation*}

where $N_{\mathrm{fixed\_rule}}$ denotes the number of design-rule violations completely eliminated after repair, and $N_{\mathrm{original\_rule}}$ denotes the initial number of detected design-rule violations before repair. This metric is referred to as the \emph{repair rate} in the subsequent analysis, consistent with the naming convention used for the preceding repair metrics.

\textbf{Decision Criteria}. A repair is considered successful if the targeted violation type is absent from the repaired diagram after re-analysis with SDMetrics. If the corresponding violation persists, the repair is considered unsuccessful.

\begin{figure}[h]
\scriptsize
\centering
\includegraphics[width=0.9\textwidth]{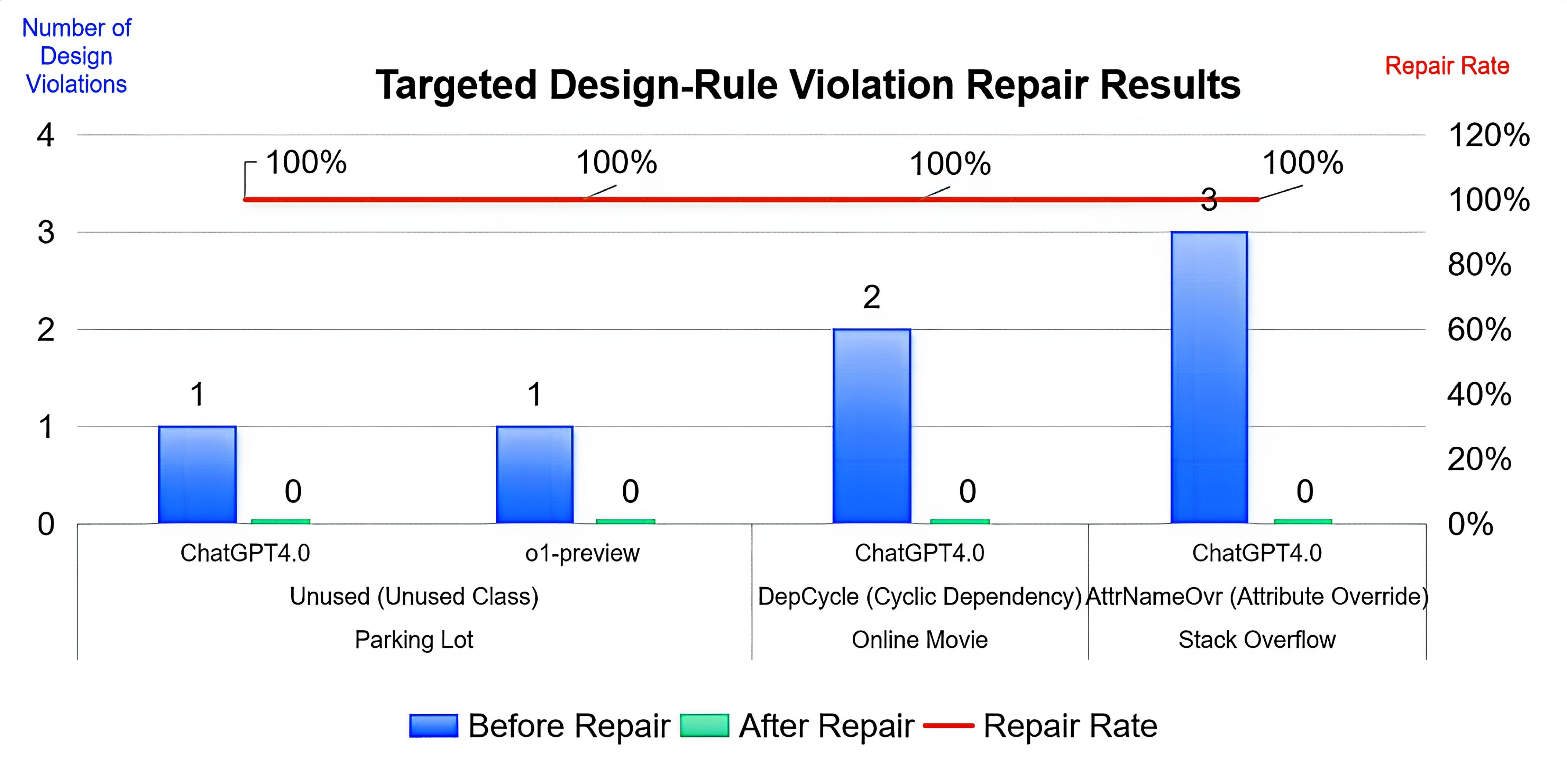}
\caption{Targeted Design-Rule Violation Repair Results}
\label{fig:scorediff4}
\end{figure}

\textbf{Quantitative Results of Design-Rule Violation Repair}. Figure~\ref{fig:scorediff4} presents the targeted repair results for design-rule violations in the class diagrams. We use the rule violation repair rate, denoted by $R_{\mathrm{rule}}$, to quantify the proportion of initially detected design-rule violations that are eliminated after repair.

In the Parking Lot case, both the o1-preview and ChatGPT-4.0-generated class diagrams contained one Unused Class violation. After remediation, both models successfully resolved the violation by integrating the functionality of the unused class into the core parking-lot class, resulting in a repair rate of $R_{\mathrm{rule}}=100\%$. This refactoring clarified the relationships among the system modules and may reduce the associated maintenance burden.

In the Online Movie case, the class diagram generated by ChatGPT-4.0 contained a DepCycle violation, indicating circular dependencies. During the repair stage, ChatGPT-4.0 resolved the two detected circular dependencies by removing the \texttt{tickets} property from the \texttt{Screening} class, thereby eliminating the bidirectional references between \texttt{Screening} and \texttt{Ticket}. This yielded a repair rate of $R_{\mathrm{rule}}=100\%$ and resulted in a less coupled and more testable design.

In the Stack Overflow case, the class diagram generated by ChatGPT-4.0 contained AttrNameOvr violations, indicating attribute-name overriding. During remediation, the \texttt{content} attribute of the parent class \texttt{Item} was renamed to \texttt{itemContent}, while the corresponding attribute of the child class \texttt{Question} was renamed to \texttt{questionContent}. This removed the naming conflicts between the parent and child classes and resolved all three detected attribute-overriding violations, resulting in $R_{\mathrm{rule}}=100\%$. The change improves naming clarity and reduces the risk of confusion during implementation and maintenance.

This repair campaign addressed three categories of critical design issues. In the Parking Lot case, the Unused Class violation was resolved by integrating the unused functionality into the core parking-lot class, thereby simplifying the model and reducing unnecessary structural elements. In the Online Movie case, two circular dependencies were eliminated, resulting in a less coupled module structure. In the Stack Overflow case, three attribute-overriding violations were corrected by clarifying the attribute names in the parent and child classes, improving naming consistency and inheritance semantics. All three repair tasks achieved a 100\% repair rate under the SDMetrics-based evaluation criteria. These results indicate improvements in structural clarity, maintainability, and conformance to the evaluated design rules, although they do not by themselves establish corresponding gains in runtime performance.

\begin{mdframed}
    \textbf{Key Findings of RQ2}: The proposed repair strategies---memory reinforcement, external knowledge injection, and detection-guided targeted repair---improve the quality of LLM-generated UML class diagrams and mitigate several common defects, including missing key classes, incorrect or omitted relationships, insufficient inheritance, unused classes, and circular dependencies. Core-class identification accuracy increased from an average of 71\% to 85\%, with improvements observed across all evaluated systems. The average repair rate for coupling relationships was 46\%. Memory reinforcement substantially reduced relationship-modeling errors for o1-preview and ChatGPT-4, whereas DeepSeek-R1 showed more modest gains and tended to transform some high-risk errors into medium-risk errors, indicating that relationship handling remains an area for improvement. Memory reinforcement also improved inheritance-group recognition and partially reduced structural complexity, with DIT and CLD decreasing from 2 to 1 in the Stack Overflow case for ChatGPT-4 and DeepSeek-R1. The inheritance relationship repair rate was 69\%. Detection-guided targeted repair eliminated 100\% of the three evaluated critical design-violation categories---Unused Class, circular dependency, and attribute overriding---thereby improving structural clarity, maintainability, and compliance with the evaluated design rules.
\end{mdframed}

\section{Discussion}\label{sec:discussion} 
In this section, we discuss the experimental results presented in Section~\ref{sec:results}, analyze the factors underlying LLM performance in UML class diagram modeling, and clarify the mechanisms and engineering value of the proposed repair strategies. We further examine the practical implications of these findings for software modeling. Together, the results provide both theoretical insights and practical guidance for the effective application of LLMs in software engineering.

\subsection{Interpretation of RQ1 Results}
The results for RQ1 indicate that differences in modeling capabilities between LLMs and human experts result in systematic deficiencies in LLM-generated UML class diagrams across four core dimensions: class identification and coverage, interclass relationship modeling, inheritance complexity control, and compliance with design rules. These modeling deficiencies are not isolated errors attributable to individual models; rather, they represent pervasive limitations observed across all evaluated LLMs, encompassing both shared inherent weaknesses and model-specific shortcomings. Fundamentally, these limitations arise from four critical disparities between LLMs and human designers: insufficient use of domain knowledge, inadequate business-logic reasoning, a lack of holistic architectural thinking, and ineffective application of object-oriented design principles in practice.

\textbf{Class identification and coverage}. As illustrated in Table~\ref{tab:llm_expert_class_coverage_table2}, LLMs predominantly rely on explicit textual cues in functional requirements to extract classes. The o1-preview model achieves the highest core-class coverage rate of 93.6\% among all evaluated models. Nevertheless, all LLMs exhibit limited coverage of implicit business entities and management-oriented classes for non-functional requirements, such as access control and system logging. The maximum coverage of implicit classes across models is only 71.4\%, while management-class coverage is extremely scarce: only o1-preview yields a non-zero coverage rate of 10\%, whereas GPT-4 and DeepSeek-R1 achieve zero coverage. Human experts leverage professional domain knowledge and practical business experience to infer implicit business rules and proactively supplement necessary management modules from a global system design perspective. In contrast, LLMs merely capture explicit information from requirement texts. Their limited domain knowledge reserve and weak business-logic reasoning hinder reliable inference beyond literal descriptions, thereby resulting in incomplete class identification. Furthermore, requirement readability exerts heterogeneous influences on different models due to distinct semantic parsing mechanisms, which further intensifies the fluctuation of class coverage performance.

\textbf{Modeling relationships between classes}. Table~\ref{tab:relationship_error_weight} demonstrates that all evaluated LLMs produce high-risk relationship errors during class diagram construction. Three error types occur most frequently (see Table~\ref{tab:error_type_semantic_matrix}): misclassifying associations as composition or aggregation (t1), misclassifying composition structures as ordinary associations (t2), and omitting essential association relationships (t4). Quantitatively, GPT-4 yields the highest average weighted risk score of 2.1, while DeepSeek-R1 delivers relatively stable relationship modeling with a lower score of 1.4. Human experts determine class relationships based on domain-specific semantic dependencies and entity lifecycle coupling, strictly distinguishing lifecycle-bound composition and loose aggregation following standardized design heuristics. However, LLMs primarily depend on lexical co-occurrence and surface-level semantic similarity. Restricted by insufficient domain knowledge and deficient business-logic reasoning, LLMs fail to perceive in-depth business constraints and lifecycle semantics. Combined with the lack of holistic architectural awareness for global relationship coordination, this limitation causes frequent relationship misclassification and critical association omissions. In addition, a notable model-scenario interaction effect is observed: modeling performance degrades substantially as the number of entities and relationship density increases, revealing the inability of LLMs to flexibly apply design rules in complex business contexts.

\textbf{Inheritance complexity control}. Table~\ref{tab:inheritance_relationship_params} reveals a polarized inheritance design pattern in LLM-generated class diagrams, which deviates from the expert-oriented principle of shallow inheritance and composition priority. In business-centric scenarios including the parking lot and online movie projects, GPT-4 largely omits valid inheritance relationships. Conversely, in the knowledge-oriented Stack Overflow scenario, both GPT-4 and DeepSeek-R1 construct overly complex inheritance hierarchies, with maximum DIT and CLD values reaching 2. Human experts balance code reusability, polymorphism, and structural simplicity through holistic architectural planning to constrain redundant inheritance depth. In comparison, LLMs struggle to balance reuse benefits and structural complexity costs due to immature practical application of software design heuristics and the absence of global system thinking. Defective business-logic reasoning further leads to two extreme behaviors: failure to identify valid generalization relationships that results in insufficient inheritance, and over-interpretation of semantic relatedness that causes inflated hierarchical structures.

\textbf{Design-rule compliance}. Table~\ref{tab:class_diagram_rule_detection} shows significant differences among LLMs in structural design-rule compliance. GPT-4 commits multiple types of violations, including unused classes, circular dependencies, and attribute overriding, resulting in a high weighted cumulative violation score of 13. The o1-preview model produces only one violation related to redundant unused classes, while DeepSeek-R1 achieves zero violations, fully matching the compliance level of expert-crafted diagrams. Human experts integrate standardized design constraints throughout the modeling process and proactively avoid structural anomalies via holistic architectural control and experienced design heuristics. In contrast, most LLMs prioritize the completeness of entity and relationship generation over overall structural rationality. Limited by insufficient domain cognition, weak business-logic constraint capability, and the lack of systematic global specification awareness, LLMs cannot consistently enforce design rules, thereby producing structurally inconsistent and non-compliant class diagrams.

\textbf{Overall assessment}. Although LLMs are capable of automatically generating UML class diagrams, their modeling outputs lack stable reliability for practical industrial deployment. The core limitations of LLM modeling precisely correspond to the four fundamental capability gaps summarized above: insufficient domain knowledge utilization, inadequate business-logic reasoning, absence of holistic architectural thinking, and ineffective implementation of object-oriented design heuristics. To bridge the performance gap between LLM-generated and expert-crafted class diagrams and mitigate these systematic modeling deficiencies, this study proposes targeted prompt optimization and expert-guided strategies. The effectiveness of these remediation approaches is comprehensively evaluated and validated in RQ2.

\subsection{Interpretation of RQ2 Results}
The experimental results for RQ2 validate the effectiveness of the three strategies proposed in this study---external knowledge injection, memory enhancement, and detection-guided targeted repair---for improving the quality of LLM-generated UML class diagrams. Each strategy effectively addresses specific modeling defects, although their impacts vary across defect types. Collectively, these strategies compensate for the modeling limitations of LLMs along three dimensions: knowledge supplementation, capability enhancement, and error correction.

\subsubsection{External Knowledge Injection}
External knowledge injection helps LLMs acquire core-class identification principles by providing domain knowledge represented through UML class diagrams, including the five principal classification criteria for entity classes. Figure~\ref{fig:scorediff1} presents the results of core-class correction after applying this strategy. Core-class coverage increased substantially, from an average of 71\% to 85\%. The underlying mechanism is to narrow the gap between the specialized software-modeling knowledge possessed by human experts and that available to the model. Specifically, expert knowledge is transformed into explicit recognition rules that the model can interpret and apply. This enables LLMs to move beyond passively extracting information from the requirements and to identify core classes in accordance with modeling principles. For example, in the online-movie scenario, both GPT-4 and o1-preview achieved 100\% core-class coverage after external knowledge injection. This result suggests that the strategy can effectively mitigate omissions arising from the models' insufficient understanding of core-class identification.

However, the strategy provides only limited improvement for implicit and management-related classes. The models continued to struggle to infer implicit business rules and to introduce classes associated with non-functional requirements. Identifying such classes requires not only general domain knowledge but also scenario-specific engineering experience. Consequently, the injection of generic rules alone is insufficient to achieve effective ``rules-plus-scenario'' customization. A promising direction for future work is therefore to develop scenario-specific knowledge injection mechanisms that are better aligned with the characteristics of individual application domains.

\subsubsection{Memory Reinforcement}
The memory reinforcement strategy is inspired by human memory processes and uses structured reasoning prompts to activate and apply UML-related knowledge. Following manual validation, a curated memory set is provided to guide the reconstruction of inter-class relationships and inheritance hierarchies. This strategy primarily targets relationship and inheritance defects, and its effectiveness varies substantially across models.

For coupling and relationship defects, Figure~\ref{fig:scorediff2} shows that o1-preview and GPT-4 achieved repair rates of 78.6\% and 76.19\%, respectively, whereas DeepSeek-R1 achieved only 26.67\%. DeepSeek-R1 also frequently transformed high-risk errors into medium-risk errors rather than eliminating them completely. The stronger performance of o1-preview and GPT-4 may be attributed to their more consistent natural-language reasoning and more reliable adherence to the prompt logic, which facilitate the retrieval and application of UML relationship definitions, notational conventions, and semantic distinctions. Although DeepSeek-R1 demonstrated strong general reasoning capabilities, its comparatively weaker UML-specific knowledge and less consistent prompt adherence may have limited its ability to distinguish among relationship types. From an architectural-quality perspective, the repaired models showed improvements of more than 50\% in maintainability, scalability, and reliability. This result suggests that memory reinforcement can alleviate architecture-level problems associated with relationship errors and produce models that are more consistent with engineering practices.

For inheritance defects, Figure~\ref{fig:scorediff3} shows that o1-preview achieved the highest repair rate at 91.67\%, followed by DeepSeek-R1 at 66.67\% and GPT-4 at 50\%. By guiding the model through a sequence of memory activation, risk bounding, diagnosis, and reconstruction, the strategy promotes the application of inheritance design principles and complexity constraints. It can therefore address both missing inheritance relationships and excessive inheritance hierarchies. For example, in the Stack Overflow scenario, DIT and CLD decreased from 2 to 1 after repair, resulting in a shallower and more appropriate hierarchy. Nevertheless, inheritance-set coverage remained insufficient in some scenarios. In the parking-lot scenario, for instance, GPT-4 achieved only 25\% inheritance-set coverage after repair, suggesting a persistent difficulty in recognizing generalization relationships among certain entities. This finding highlights the need to supplement memory reinforcement with scenario-specific business knowledge.

Overall, memory reinforcement helps reduce ambiguity in the application of UML knowledge by guiding LLMs from loosely associated reasoning toward more explicit and structured modeling decisions. Its effectiveness remains highly dependent on each model's reasoning capability, UML-specific knowledge, and ability to follow the provided prompts, demonstrating clear model-specific variation.

\subsubsection{Targeted Repair}
The targeted repair strategy uses SDMetrics detection results to address three categories of design-rule violations---unused classes, circular dependencies, and attribute overriding---through tailored prompts. As shown in Figure~\ref{fig:scorediff4}, it achieved a 100\% repair rate. The strategy combines automated error localization with LLM-based model reconstruction: SDMetrics identifies the location and type of each violation, after which the LLM performs focused modifications to resolve the detected problem. This process enables precise repairs while minimizing unnecessary changes to the overall model.

The repairs produced several concrete outcomes. In the parking-lot scenario, unused classes were consolidated into core classes, thereby reducing structural redundancy. In the online-movie scenario, circular dependencies were eliminated by removing redundant attributes and breaking bidirectional references. In the Stack Overflow scenario, semantic consistency within the inheritance hierarchy was restored by renaming conflicting attributes in the parent and child classes. These modifications removed the evaluated rule violations and improved structural clarity, readability, and maintainability, bringing the generated diagrams into full conformance with the evaluated design rules.

The effectiveness of targeted repair can be attributed largely to its ability to avoid the weaknesses of fully autonomous, end-to-end redesign. Automated tools provide objective error localization, while the LLM is constrained to perform localized corrections rather than reconstructing the entire model. This division of responsibilities improves the reliability and controllability of the repair process. More broadly, the findings support human--AI collaboration as an effective paradigm for LLM-assisted software modeling: human experts can oversee global design decisions and quality assurance, while LLMs perform localized modifications and repetitive correction tasks, allowing both parties to contribute their complementary strengths.

\subsection{Implications}
The findings for RQ1 and RQ2 yield several practical implications for the deployment of LLMs in software modeling. The central implication is the value of a human--AI collaborative modeling paradigm that combines the generative efficiency of LLMs with the engineering judgment of human experts to improve both productivity and model quality.

\textbf{Model-specific tailoring}: Modeling and repair strategies should be adapted to the capabilities and limitations of individual models. The experiments revealed substantial heterogeneity in both modeling performance and repair effectiveness. o1-preview performed best in core-class identification and achieved strong results in relationship and inheritance repair. DeepSeek-R1 performed well in design-rule compliance, whereas GPT-4 showed considerable repair potential in selected scenarios. Model selection should therefore be guided by project requirements. For example, o1-preview may be preferable when accurate core-class identification is the primary objective, while DeepSeek-R1 may be more suitable when strict design-rule conformance is emphasized. In either case, model-specific repair strategies should be used to address the weaknesses of the selected model and improve model--strategy alignment.

\textbf{Closed-loop toolchain integration}: Automated detection tools should be integrated closely with LLMs to establish a closed-loop modeling workflow. In this study, SDMetrics was used to assess the quality of UML class diagrams and was combined with LLM-based repair to implement a generate--detect--repair--validate cycle. In practice, this workflow could be operationalized by integrating SDMetrics, PlantUML, Visual Paradigm, and LLMs into a unified intelligent modeling platform. Such a platform could support automated diagram generation, quality assessment, and defect correction, thereby improving modeling efficiency and process consistency.

\textbf{Prompt-engineering optimization}: Prompt engineering should be treated as a central mechanism for improving LLM-based software modeling. All three repair strategies were prompt-driven, including chain-of-thought prompts, domain-knowledge prompts, and targeted repair prompts. Practitioners should therefore design structured and fine-grained prompts that are aligned with UML standards, software design principles, and scenario-specific characteristics. Such prompts can provide clearer constraints and guide models toward class diagrams that are more consistent with the requirements and the intended design rules.

\textbf{Expert-crafted oversight}: Human experts should retain a central oversight role throughout the modeling process. Although the proposed repair strategies substantially improved the LLM outputs, the models could not fully replace expert judgment. They continued to face difficulties in identifying implicit and management-related classes, modeling relationships in complex scenarios, and maintaining a holistic architectural perspective. Within a human--AI collaborative workflow, experts should lead requirements analysis, global architectural design, and final quality assurance, while LLMs should support initial drafting, localized editing, and repetitive correction tasks. This division of labor can reduce the routine workload of experts while allowing them to focus on high-value design decisions.

\section{Threats to Validity}\label{sec:Threats} 
This study conducted systematic experiments to assess the quality of LLM-generated UML class diagrams and evaluate corresponding repair strategies. Despite the use of a rigorous research design and multifaceted validation procedures, the findings may still be affected by methodological choices, experimental data, and model-specific characteristics. Following established frameworks for assessing threats to validity in empirical software engineering, this section examines construct validity, internal validity, external validity, and conclusion validity, identifies potential risks, and discusses the mitigation measures adopted, together with directions for future improvement.

\textbf{Construct validity}: Three primary threats warrant consideration. (1) The inherent stochasticity of LLMs means that identical requirements and prompts may produce different diagrams. To mitigate this issue, we generated three candidate diagrams for each case and selected the best-performing candidate. Although this procedure reduces variance attributable to randomness, it does not eliminate it entirely. (2) The evaluation constructs are not fully comprehensive. Our framework covers scope, relationships, structural complexity, and design-rule conformance, and combines SDMetrics with human review. However, some aspects---such as the appropriateness of implicit-class identification and the necessity of management-related classes---remain dependent on subjective judgment. In addition, engineering properties such as extensibility and testability were not quantified. (3) The quantification of relationship errors relies on a semantic error-type matrix whose weights were derived from the research team's engineering experience and impact analysis rather than from industry-wide standards. This choice may introduce bias into the estimated risk values. Mitigation measures included the use of a standardized experimental protocol, multiple rounds of review with inter-rater consistency checks, and expert elicitation to determine the weights. Future work will align the evaluation metrics more closely with industry standards, introduce quantitative indicators for extensibility and testability, refine relationship-error weights through larger expert surveys, and use multi-sample averaging to further reduce the influence of generative randomness.

\textbf{Internal validity}: Three main sources of threat were identified. (1) Incomplete control of generation parameters: although prompts, outputs, and experimental procedures were harmonized, parameters such as temperature and maximum output length were not fully standardized across models. These differences may have confounded the comparison. (2) Human evaluation bias: the assessment of core-class identification and inheritance reasonableness relied partly on human reviewers. Although we employed double-blind assessment and calculated Fleiss' kappa to measure inter-rater agreement, differences in the reviewers' software engineering backgrounds and UML expertise may still have introduced subjectivity. (3) Toolchain-related discrepancies: the use of SDMetrics, Visual Paradigm, and PlantUML may have introduced minor inconsistencies because of differences in UML semantic parsing and format conversion, thereby affecting metric precision. To mitigate these risks, we kept generation parameters as consistent as possible, provided reviewers with standardized training and evaluation guidelines, and cross-validated the results across tools.

\textbf{External validity}: External validity concerns the extent to which the findings generalize across scenarios, models, datasets, and practical development contexts. Three considerations are particularly important. (1) Case scope and domain coverage: the study used three cross-domain cases---parking lot, online movie, and Stack Overflow---representing IoT, e-commerce, and knowledge-service contexts, respectively. The systems were relatively small or medium-sized, containing 13--28 classes. These controlled cases provide a useful testbed for evaluating UML class modeling and repair workflows, but they may not capture the complexity of large-scale systems. Future evaluations involving large distributed systems, embedded systems, and other complex engineering contexts will help broaden the applicability of the findings. (2) Model breadth: the experiments included three representative LLMs---GPT-4, o1-preview, and DeepSeek-R1---which provide a comparative baseline. Future work should expand the evaluation to additional model families and lightweight open-source LLMs to improve cross-model coverage. (3) Engineering realism: the requirements were standardized to ensure comparability and isolate the effects of modeling and repair strategies. Subsequent studies should incorporate industrial factors, such as evolving requirements and multi-team collaboration, and validate the proposed approach in real-world projects to further assess its generalizability. (4)  Reliance on Expert Guidance and Framework Deployment Scalability:
The observed repair rates for coupling relationships, inheritance structures, and core‑class identification primarily demonstrate the performance of our proposed general‑purpose prompting‑guidance framework, rather than the intrinsic self‑repair capabilities of vanilla LLMs. All reasoning chains, prompt templates, and defect‑repair rules adopted in the experiments were systematically designed and reviewed by UML experts. Unlike case‑specific prompts, the proposed prompting‑guidance framework is designed to be general and reusable for UML class‑diagram modeling tasks and is not exclusively customized for any single experimental case. Nevertheless, constructing and configuring this guidance framework still demands specialized UML domain knowledge and manual expert intervention. Although the framework itself supports general‑case reuse, such human involvement introduces extra application overhead and constrains its scalability and efficiency for large‑scale industrial deployment. This reliance on expert input therefore constitutes an important limitation of the present approach.

\textbf{Conclusion validity}: Conclusion validity concerns the statistical robustness and reliability of the findings. Within the scope of the study, the conclusions are supported by multidimensional evaluation and repeated trials. Nevertheless, three aspects could further strengthen the validity of the conclusions. (1) Sample scope: the dataset comprised three cases, and each LLM contributed one best-performing class diagram per case. Although this design revealed consistent patterns across heterogeneous domains, future work should expand the case pool and increase the number of generated samples per model and case. (2) Statistical analysis: effectiveness was evaluated using descriptive indicators, including coverage, repair rate, and weighted error scores, which are appropriate for the mechanism-oriented focus of this study. Subsequent research should complement these measures with inferential analyses, such as hypothesis testing and analysis of variance, to quantify statistical significance and estimate uncertainty. (3) Temporal horizon: the evaluation focused on immediate post-repair quality in order to isolate the effects of the proposed strategies. Follow-up experiments should examine downstream effects on code generation, maintenance, and iterative development, thereby enabling a more comprehensive assessment of engineering value across the software lifecycle.

\section{Conclusions and Future Work}\label{conclusions}
This study presents an exploratory investigation into the quality evaluation and defect remediation of LLM-generated UML class diagrams. Using three cross-domain cases and three representative LLMs---GPT-4, o1-preview, and DeepSeek-R1---we developed a multidimensional evaluation framework covering size and completeness, relationship correctness, inheritance hierarchy, and design rule compliance. By combining automated quantification with SDMetrics and expert manual review, the study compares the UML modeling performance of LLMs with that of professional experts in the evaluated scenarios. We also propose and empirically assess three targeted remediation strategies: external knowledge injection, memory reinforcement, and detection-guided targeted repair.

The experimental results reveal several recurring defects in LLM-generated UML class diagrams, including incomplete identification of core classes, incorrect relationship modeling, inappropriate inheritance structures, and violations of design rules. These findings indicate that, in the evaluated scenarios, LLMs produced models that were generally inferior to expert-crafted benchmarks. Within the scope of the experiments, the proposed remediation strategies achieved measurable improvements across all evaluated dimensions. In particular, the repair rate for design-rule violations reached 100\%, while the average correction rate for core-class identification reached 85\%. These results provide preliminary evidence of the effectiveness and practical potential of the proposed strategies in the tested scenarios. More broadly, the findings suggest that human--AI collaboration can help compensate for the current limitations of LLMs in software modeling by combining their generative efficiency with expert knowledge and quality oversight.

Future work will extend and strengthen this exploratory investigation in several directions. (1) We will integrate domain ontologies and engineering-experience repositories into the repair process to improve LLMs' ability to identify implicit and management-related classes. (2) We will expand the scale and complexity of the cases, as well as the range of evaluated models, to improve the generalizability of the findings. (3) We will extend the research to other UML diagram types, such as use case and sequence diagrams, with the goal of developing a more comprehensive quality-assurance framework for intelligent UML modeling. (4) We will investigate the engineering implementation of the proposed methods by developing an integrated intelligent modeling platform that supports automated requirements analysis, diagram generation, defect detection and repair, and result export. Such a platform would facilitate further validation of the approach in realistic software-development settings.  

\section*{Data availability}
The replication package for this work has been made available at~\citep{replpack}.


\section*{Acknowledgements}
This work has been partially supported by the National Natural Science Foundation of China (NSFC) with Grant No. 92582203.

\section*{CRediT authorship contribution statement}
\textbf{Jie Liang:} Conceptualization, Investigation, Data curation, Formal analysis, Writing - Original draft preparation. \textbf{Peng Liang:} Conceptualization, Methodology, Investigation, Data curation, Supervision, Writing - Original draft preparation. \textbf{Chong Wang:} Conceptualization, Methodology, Writing - review and editing.


\bibliography{references}

@article{camara2024towards,
  title={Towards standardized benchmarks of {LLM}s in software modeling tasks: a conceptual framework},
  author={C{\'a}mara, Javier and Burgue{\~n}o, Loli and Troya, Javier},
  journal={Software and Systems Modeling},
  volume={23},
  number={6},
  pages={1--10},
  year={2024},
  publisher={Springer}

}

@article{hou2024large,
  title={Large Language Models for Software Engineering: A Systematic Literature Review},
  author={Hou, Xinyi and Zhao, Yanjie and Liu, Yue and Yang, Zhou and Wang, Kailong and Li, Li and Luo, Xiapu and Lo, David and Grundy, John and Wang, Haoyu},
  journal={ACM Transactions on Software Engineering and Methodology},
  volume={33},
  number={8},
  pages={1--79},
  year={2024},
  publisher={ACM}
}

@inproceedings{fan2023large,
  title={Large Language Models for Software Engineering: Survey and Open Problems},
  author={Fan, Angela and Gokkaya, Beliz and Harman, Mark and Lyubarskiy, Mitya and Sengupta, Shubho and Yoo, Shin and Zhang, Jie M},
  booktitle={Proceedings of the 45th IEEE/ACM International Conference on Software Engineering: Future of Software Engineering (ICSE-FoSE)},
  pages={31--53},
  year={2023},
  publisher={IEEE}
}

@inproceedings{ahmad2023human,
  title={Towards Human-Bot Collaborative Software Architecting with {ChatGPT}},
  author={Ahmad, Aakash and Waseem, Muhammad and Liang, Peng and Fahmideh, Mahdi and Aktar, Md Saiful and Mikkonen, Tommi},
  booktitle={Proceedings of the 27th International Conference on Evaluation and Assessment in Software Engineering (EASE '23)},
  pages={279--285},
  year={2023},
  publisher={ACM}
}

@article{marques2024chatgpt,
  title={Using ChatGPT in Software Requirements Engineering: A Comprehensive Review},
  author={Marques, Nuno and Silva, Rodrigo Rocha and Bernardino, Jorge},
  journal={Future Internet},
  volume={16},
  number={6},
  pages={1--21},
  year={2024},
  publisher={MDPI}
}

@article{naveed2024mde,
  title={Model Driven Engineering for Machine Learning Components: A Systematic Literature Review},
  author={Naveed, Hussain and Arora, Chetan and Khalajzadeh, Hourieh and Grundy, John and Haggag, Omar},
  journal={Information and Software Technology},
  volume={169},
  pages={107423},
  year={2024},
  publisher={Elsevier}
}

@inproceedings{disipio2024llms,
  title={On the use of LLMs to support the development of domain-specific modeling languages},
  author={Di Rocco, Juri and Di Ruscio, Davide and Di Sipio, Claudio and Nguyen, Phuong T. and Rubei, Riccardo},
  booktitle={Proceedings of the 27th ACM/IEEE International Conference on Model Driven Engineering Languages and Systems (MODELS): Companion},
  pages={596--601},
  year={2024},
  publisher={ACM}
}

@misc{wust2005sdmetrics,
  author = {Wüst, Jürgen},
  title = {{SDMetrics}: The Software Design Metrics Tool for {UML}},
  howpublished = {\url{https://www.sdmetrics.com/}},
  year = {2005}
}

@inproceedings{siala2025using,
  title={Using {LLMs} to extract {UML} class diagrams from {Java} and {Python} programs: an empirical study},
  author={Siala, H. A. and Lano, K.},
  booktitle={Proceedings of the STAF 2025 Workshops: Agile Model-Driven Engineering (AMDE)},
  pages={1--10},
  year={2025},
  publisher={CEUR-WS.org}
}

@article{jiang2025survey,
  title={A Survey on Large Language Models for Code Generation},
  author={Jiang, Juyong and Wang, Fan and Shen, Jiasi and Kim, Sungju and Kim, Sunghun},
  journal={ACM Transactions on Software Engineering and Methodology},
  volume={35},
  number={2},
  pages={1--72},
  year={2025},
  publisher={ACM}

}

@inproceedings{wang2024llms,
  title={How LLMs Aid in {UML} Modeling: An Exploratory Study with Novice Analysts},
  author={Wang, Beian and Wang, Chong and Liang, Peng and Li, Bing and Zeng, Cheng},
  booktitle={Proceedings of the 21st IEEE International Conference on Software Services Engineering (SSE)},
  pages={249--257},
  year={2024},
  publisher={IEEE}
}

@article{camara2023assessment,
  title={On the assessment of generative {AI} in modeling tasks: an experience report with {ChatGPT} and {UML}},
  author={C{\'a}mara, Javier and Troya, Javier and Burgue{\~n}o, Loli and Vallecillo, Antonio},
  journal={Software and Systems Modeling},
  volume={22},
  number={3},
  pages={781--793},
  year={2023},
  publisher={Springer}
}

@inproceedings{debari2024evaluating,
  title={Evaluating Large Language Models in Exercises of {UML} Class Diagram Modeling},
  author={De Bari, Dario and Garaccione, Giulia and Coppola, Riccardo and Ardito, Luca and Torchiano, Marco},
  booktitle={Proceedings of the 18th ACM/IEEE International Symposium on Empirical Software Engineering and Measurement (ESEM)},
  pages={393--399},
  year={2024},
  publisher={ACM}
}

@inproceedings{shehata2024creating,
  title={Creating UML class diagrams with general-purpose {LLMs}},
  author={Shehata, Mina and Lepore, Blaire and Cummings, Hailey and Parra Rodriguez, Esteban},
  booktitle={Proceedings of the 20th IEEE Working Conference on Software Visualization (VISSOFT)},
  pages={157--158},
  year={2024},
  publisher={IEEE}
}

@article{joel2024survey,
  title={A Survey on LLM-Based Code Generation for Low-Resource and Domain-Specific Programming Languages},
  author={Joel, Sathvik and Wu, Jie JW and Fard, Fatemeh H},
  journal={ACM Transactions on Software Engineering and Methodology},
  volume={34},
  number={6},
  pages={1--45},
  year={2025},
  publisher={ACM}
}

@article{zhang2024systematic,
  title={A systematic literature review on large language models for automated program repair},
  author={Zhang, Quanjun and Fang, Chunrong and Xie, Yang and Ma, YuXiang and Sun, Weisong and Yang, Yun and Chen, Zhenyu},
  journal={ACM Transactions on Software Engineering and Methodology, (2026)},
  year={2026},
  publisher={ACM}
}

@article{chen2024model,
  title={A Model Is Not Built By A Single Prompt: {LLM}-Based Domain Modeling With Question Decomposition},
  author={Chen, Ru and Shen, Jingwei and He, Xiao},
  journal={arXiv preprint arXiv:2410.09854, (2024)},
  year={2024}
}

@inproceedings{ferrari2024model,
  title={Model Generation with {LLM}s: From Requirements to {UML} Sequence Diagrams},
  author={Ferrari, Alessio and Abualhaija, Sallam and Arora, Chetan},
  booktitle={Proceedings of the 32nd IEEE International Requirements Engineering Conference Workshops (REW)},
  pages={291--300},
  year={2024},
  publisher={IEEE}
}

@article{feng2025integrating,
  title={Integrating Various Software Artifacts for Better {LLM}-Based Bug Localization and Program Repair},
  author={Feng, Qiong and Ma, Xiaotian and Sheng, Jiayi and Feng, Ziyuan and Song, Wei and Liang, Peng},
  journal={ACM Transactions on Software Engineering and Methodology, (2026)},
  year={2025},
  publisher={ACM}
}

@article{giannouris2026nomad,
   title={{NOMAD}: A multi-agent {LLM} system for {UML} class diagram generation from natural language requirements},
  author={Giannouris, P. and Ananiadou, S.},
  journal={arXiv preprint arXiv:2511.2240, (2025)},
  year={2025}
}

@inproceedings{jain2024livecodebench,
  title={LiveCodeBench: Holistic and Contamination Free Evaluation of Large Language Models for Code},
  author={Jain, N. and Han, K. and Gu, A. and Li, W.-D. and Yan, F. and Zhang, T. and Wang, S. and Solar-Lezama, A. and Sen, K. and Stoica, I.},
  booktitle={Proceedings of the 13th International Conference on Learning Representations (ICLR)},
  year={2025},
  pages={1--41},
  publisher={OpenReview}
}

@article{kusmaryono2022number,
  title={Number of response options, reliability, validity, and potential bias in the use of the likert scale education and social science research: A literature review},
  author={Kusmaryono, Imam and Wijayanti, Dyana and Maharani, Hevy Risqi},
  journal={International Journal of Educational Methodology},
  volume={8},
  number={4},
  pages={625--637},
  year={2022},
  publisher={Eurasian Society of Educational Research}
}

@misc{manualdesigns2025parking,
  title={Parking Lot System Design},
  author={{Manual Designs}},
  howpublished={\url{https://rajat19.github.io/system-design/system-designs/parking-lot.html}},
  year={2025},
  note={Accessed: 2026-03-15}
}

@misc{manualdesigns2025movie,
  title={Movie Booking System Design},
  author={{Manual Designs}},
  howpublished={\url{https://rajat19.github.io/system-design/system-designs/movie-booking.html}},
  year={2025},
  note={Accessed: 2026-03-15}
}

@misc{manualdesigns2025stack,
  title={Stack Overflow System Design},
  author={{Manual Designs}},
  howpublished={\url{https://rajat19.github.io/system-design/system-designs/stack-overflow.html}},
  year={2025},
  note={Accessed: 2026-03-15}
}

@inproceedings{li2025prompting,
  title={Prompting Large Language Models to Tackle the Full Software Development Lifecycle: A Case Study},
  author={Li, Bowen and Wu, Wenhan and Tang, Ziwei and Shi, Lin and Yang, John and Li, Jinyang and Yao, Shunyu and Qian, Chen and Hui, Binyuan and Zhang, Qicheng and Yu, Zhiyin and Du, He and Yang, Ping and Lin, Dahua and Peng, Chao and Chen, Kai},
  booktitle={Proceedings of the 31st International Conference on Computational Linguistics (COLING)},
  pages={7511--7531},
  year={2025},
  publisher={ACL},
}

@article{briand1999unified,
  title={A Unified Framework for Coupling Measurement in Object-Oriented Systems},
  author={Briand, Lionel C. and Daly, John W. and W{\"u}st, J{\"u}rgen K.},
  journal={IEEE Transactions on Software Engineering},
  volume={25},
  number={1},
  pages={91--121},
  year={1999},
  month={jan},
  publisher={IEEE}
}

@book{fenton2014software,
  title={Software Metrics: A Rigorous and Practical Approach, Third Edition},
  author={Fenton, Norman E. and Bieman, James},
  publisher={CRC Press},
  year={2014},
  month={sep},
  isbn={978-1-4398-3822-8}
}

@article{shaikh2021more,
  title={More Than Two Decades of Research on Verification of {UML} Class Models: A Systematic Literature Review},
  author={Shaikh, Asadullah and Hafeez, Abdul and Wagan, Asif Ali and Alrizq, Mesfer and Alghamdi, Abdullah and Reshan, Mana Al},
  journal={IEEE Access},
  volume={9},
  pages={142461--142474},
  year={2021},
  publisher={IEEE}
}

@inproceedings{ren2024misuse,
  title={From misuse to mastery: Enhancing code generation with knowledge-driven {AI} chaining},
  author={Ren, X. and Ye, X. and Zhao, D. and Xing, Z. and Yang, X.},
  booktitle={Proceedings of the 38th IEEE/ACM International Conference on Automated Software Engineering (ASE)},
  pages={976--987},
  year={2023},
  publisher={IEEE}
}

@book{wille2017automated,
  title={Automated Validation and Verification of {UML}/{OCL} Models Using Satisfiability Solvers},
  author={Wille, Robert and Przigoda, Nils and Przigoda, Judith and Drechsler, Rolf},
  publisher={Springer},
  year={2018}
}

@article{schacter1999seven,
  title={The Seven Sins of Memory: Insights from Psychology and Cognitive Neuroscience},
  author={Schacter, Daniel L.},
  journal={American Psychologist},
  volume={54},
  number={3},
  pages={182--203},
  year={1999},
  month={mar},
  publisher={American Psychological Association}
}

@article{richards2023persistence,
  title={The Persistence and Transience of Memory},
  author={Richards, Blake A. and Frankland, Paul W.},
  journal={Neuron},
  volume={111},
  number={10},
  pages={1531--1544},
  year={2017},
  publisher={Cell Press}
}

@inproceedings{sun2024iterative,
  title={Enhancing Chain-of-Thoughts Prompting with Iterative Bootstrapping in Large Language Models},
  author={Sun, Jiashuo and Luo, Yi and Gong, Yeyun and Lin, Chen and Shen, Yelong and Guo, Jian and Duan, Nan},
  booktitle={Proceedings of the 39th Annual Conference of the Nations of the Americas Chapter of the Association for Computational Linguistics (NAACL): Findings},
  pages={4074--4101},
  year={2024},
  publisher={ACL}
}

@article{zhou2024bias,
  title={Bias and Fairness in Large Language Models: A Survey},
  author={Gallegos, I. O. and Rossi, R. A. and Barrow, J. and Tanjim, M. M. and Kim, S. and Dernoncourt, F. and Yu, T. and Zhang, R. and Ahmed, N. K.},
  journal={Computational Linguistics},
  volume={50},
  number={3},
  pages={1097--1199},
  year={2024},
  month={sep},
  publisher={MIT Press}
}

@article{ibanez2025multimodal,
  title={Can Multimodal Large Language Models Grade Like an Expert? A Study on {UML} Class Diagram Assessment Accuracy},
  author={Ib\'a\~nez, Mar\'ia Blanca and Barr\'on-Estrada, Mar\'ia Luc\'ia and Zatarain-Cabada, Ram\'on},
  journal={Computer Applications in Engineering Education},
  volume={33},
  number={5},
  pages={e70080},
  year={2025},
  publisher={Wiley}
}

@manual{sdmetricsmanual,
  title={{SDMetrics} User Manual: The Quality Measurement Tool for {UML} Designs},
  author={{SDMetrics}},
  organization={SDMetrics},
  year={2024},
  note={Version 2.5, Accessed: 2026-03-15}
}

@article{solid2025empirical,
  title={Are We {SOLID} Yet? An Empirical Study on Prompting {LLM}s to Detect Design Principle Violations},
  author={Pehlivan, Fatih and Erg{\"u}zen, Arcin {\"U}lk{\"u} and Moslemi Yengejeh, Sahand and Lami, Mayasah and Koyuncu, Anil},
  journal={arXiv preprint arXiv:2509.03093, (2025)},
  year={2025}
}

@article{llmrepair2025survey,
  title={A Survey of {LLM}-based Automated Program Repair: Taxonomies, Design Paradigms, and Applications},
  author={Yang, Boyang and Cai, Zijian and Liu, Fengling and Le, Bach and Zhang, Lingming and Bissyand{\'e}, Tegawend{\'e} F. and Liu, Yang and Tian, Haoye},
  journal={arXiv preprint arXiv:2506.23749, (2025)},
  year={2025}
}

@article{locobench2025benchmark,
  title={A Benchmark for Long-Context Large Language Models in Complex Software Engineering},
  author={Qiu, Jielin and Liu, Zuxin and Liu, Zhiwei and Murthy, Rithesh and Zhang, Jianguo and Chen, Haolin and Wang, Shiyu and Zhu, Ming and Yang, Liangwei and Tan, Juntao and Cen, Zhepeng and Qian, Cheng and Heinecke, Shelby and Yao, Weiran and Savarese, Silvio and Xiong, Caiming and Wang, Huan},
  journal={arXiv preprint arXiv:2509.09614, (2025)},
  year={2025}
}

@misc{replpack,
  author       = {Jie Liang and Peng Liang and Chong Wang},
  title        = {{Replication Package for the Paper: UML Class Diagram Evaluation and Repair Strategies based on LLMs}},
  year         = 2026,
  publisher    = {Github},
  note         = {\url{https://github.com/JieLiang1030/UML-Class-Diagram-Evaluation}}
}

\end{document}